\documentclass[11pt]{article}
\usepackage{times}
\usepackage{amstext,amsmath,amsfonts}
\usepackage{graphicx}
\usepackage{color}
\usepackage{fancyhdr}
\usepackage{wrapfig}
\usepackage{amssymb}
\usepackage{amsthm}
\usepackage{bm}
\usepackage{setspace}                  
\usepackage{subfigure}
\fancypagestyle{plain0}{
  \fancyhf{}
  \fancyhead[L]{\footnotesize \sf 
  M. Ercsey-Ravasz and Z. Toroczkai  }   \fancyhead[R]{\footnotesize \sf Optimization hardness as transient chaos ... }
  
  \fancyfoot[C]{\thepage}
}

\newcommand{\nid}{\noindent}
\newcommand{\ms}{\medskip}

\newcommand{\bq}{\begin{equation}}
\newcommand{\eq}{\end{equation}}
\newcommand{\bqa}{\begin{eqnarray}}
\newcommand{\eqa}{\end{eqnarray}}

\begin{document}

\title{
 Optimization hardness as transient chaos in an analog approach to 
constraint satisfaction\thanks{The article appeared in {\it Nature Physics} {\bf 7}, 966 \, (2011) }}

\author{{\bf M\'aria Ercsey-Ravasz}$^{1,3}$\thanks{E-mail: ercsey.ravasz@phys.ubbcluj.ro}  \, and {\bf Zolt\'an Toroczkai}$^{1,2}$\thanks{E-mail: toro@nd.edu}\\
{$\mbox{\;\;\;}$}\\
{ $^{1}$  Department of Physics, University of Notre Dame, Notre Dame, IN, 46556 USA  and} \\
{Interdisciplinary Center for 
Network Science and Applications (iCeNSA) } \\
{ $^{2}$ Departments of Computer Science and Engineering,} \\
{ University of Notre Dame, Notre Dame, IN, 46556 USA}\\
{ $^{3}$ Faculty of Physics, 
 Babes-Bolyai University, Cluj-Napoca, Romania}}

\maketitle

\begin{abstract}
Boolean satisfiability \cite{Ref1} ($k$-SAT) is one of the most studied optimization problems, as an efficient (that is, polynomial-time) solution to $k$-SAT (for $k\geq 3$) implies efficient solutions to a large number of hard optimization problems \cite{Ref2,Ref3}. Here we propose a mapping of $k$-SAT into a deterministic continuous-time dynamical system with a unique correspondence between its attractors and the $k$-SAT solution clusters. We show that beyond a constraint density threshold, the analog trajectories become transiently chaotic \cite{Ref4,Ref5,Ref6,Ref7}, and the boundaries between the basins of attraction \cite{Ref8} of the solution clusters become fractal \cite{Ref7,Ref8,Ref9}, signaling the appearance of optimization hardness \cite{Ref10}. Analytical arguments and simulations indicate that the system always finds solutions for satisfiable formulae even in the frozen regimes of random $3$-SAT \cite{Ref11} and of locked occupation problems \cite{Ref12} (considered among the hardest algorithmic benchmarks); a property partly due to the systemÕs hyperbolic \cite{Ref4,Ref13} character.  The system finds solutions in polynomial continuous-time, however, at the expense of exponential fluctuations in its energy function.
\end{abstract}

\newpage
Boolean satisfiability \cite{Ref1}  ($k$-SAT, $k\geq 3$) is the quintessential constraint satisfaction problem, lying at the basis of many decision, scheduling, error-correction and bio-computational applications. $k$-SAT is in NP, that is its solutions are efficiently (polynomial time) checkable, but no efficient (polynomial time) algorithms are known to compute those solutions \cite{Ref2}. If such algorithms would be found for $k$-SAT, all NP problems would be efficiently computable, since $k$-SAT is NP-complete \cite{Ref2,Ref3}.

In k-SAT there are given $N$ Boolean variables $\{x_1,\dots,x_N\}$, $x_i\in \{0,1\}$   and $M$ clauses (constraints), each clause being the disjunction (OR, denoted as $\vee$) of $k$ variables or their negation ($\overline{x}$). One has to find an assignment of the variables such that all clauses (called collectively as a formula) are satisfied (TRUE = Ò1Ó).  When the number of constraints is small, it is easy to find solutions, while for too many constraints it is easy to decide that the formula is unsatisfiable (UNSAT). Deciding satisfiability, in the 'intermediate range', however, can be very hard: the worst-case complexity of all known algorithms for $k$-SAT is exponential in $N$. 

Inspired by the mechanisms of information processing in biological systems, analog computing received increasing interest from both theoretical \cite{Ref14,Ref15,Ref16} and engineering communities \cite{Ref17,Ref18,Ref19,Ref20,Ref21}. Although the theoretical possibility of efficient computation via chaotic dynamical systems has been shown previously \cite{Ref15}, nonlinear dynamical systems theory has not been exploited for NP-complete problems in spite of the fact that, as shown by Gu et al.\cite{Ref19}, Nagamatu et al. \cite{Ref20} and Wah et al. \cite{Ref21}, $k$-SAT can be formulated as a continuous global optimization problem\cite{Ref19}, and even cast as an analog dynamical system \cite{Ref20,Ref21}.

Here we present a novel continuous-time dynamical system for $k$-SAT, with a dynamics that is rather different from previous approaches. Let us introduce the continuous variables \cite{Ref19} $s_i\in[-1,1]$ , such that $s_i=-1$ if the $i$-th variable ($x_i$) is FALSE and $s_i=1$  if it is TRUE.  We define $c_{mi}=1$  for the direct form ($x_i$), $c_{mi}=-1$  for the negated form ($\overline{x}_i$), and $c_{mi}=0$  for the absence of the $i$-th variable from clause $m$. Defining the constraint function $K_m(\bm{s})\equiv 2^{-k}\prod_{i=1}^N (1-c_{mi}s_i)$   corresponding to clause $m$, we have $K_m\in[0,1]$  and $K_m=0$   if and only if clause $m$ is satisfied. The goal would be to find a solution $\bm{s^*}$ with $s_i^*\in\{-1,1\}$  to $E(\bm{s^*})=0$, where $E$ is the energy function $E(\bm{s})=\sum_{m=1}^M K_m(\bm{s})^2$. If such $\bm{s^*}$ exists, it will be a global minimum for $E$ and a solution to the $k$-SAT problem. However, finding $\bm{s^*}$ by a direct minimization of $E(\bm{s})$   will typically fail due to non-solution attractors trapping the search dynamics.  In order to avoid such traps, here we define a modified energy function 
$V(\bm{s,a})=\sum_{m=1}^M a_mK_m(\bm{s})^2$, using auxiliary variables $a_m\in(0,\infty)$  similar to Lagrange multipliers \cite{Ref20,Ref21}.  Let us denote by ${\cal H}_N$ the continuous domain $[-1,1]^N$. Its boundary is the $N$-hypercube $Q_N=\partial{\cal H}_N$ with vertex set 
${\cal V}_N=\{-1,1\}^N \subset Q_N$. The set of solutions for a given $k$-SAT formula, called solution space, occupies a subset of ${\cal V}_N$. {\it Solution clusters} are formed by solutions that can be connected via {\it single-variable} flips, always staying within satisfying assignments \cite{Ref22}. Clearly, $V\geq0$  in $\Omega\equiv{\cal H}_N\times (0,\infty)^M$, and $V(\bm{s,a})=0$ {\it within} ${\cal V}_N$   if and only if $\bm{s}=\bm{s^*}\in{\cal V}_N$  is a $k$-SAT solution, for any $\bm{a}\in(0,\infty)^M$. We now introduce a {\it continuous-time dynamical system} on $\Omega$ through: 
\begin{eqnarray}
&&\frac{ds_i}{dt} = \left(- \nabla_s V(\bm{s},\bm{a})\right)_i= \sum_{m=1}^{M} 
2 a_m c_{mi} K_{mi}(\bm{s}) K_m(\bm{s})\;,\;\;\;\;i=1,\ldots,N\;, \label{1a}  \\
&& \frac{da_m}{dt} =  a_m K_m(\bm{s})\;,\;\;\;\;m=1,\ldots,M\;, \label{1b}
\end{eqnarray}
where $\nabla_s$   is the gradient operator with respect to $\bm{s}$,  and $K_{mi}=K_m/(1-c_{mi}s_i)$. The initial conditions for $\bm{s}$ are arbitrary  $\bm{s}(0)\in{\cal H}_N$, however, for $\bm{a}$  they have to be strictly positive, $a_m(0)>0$ (e.g., $a_m(0)=1$). The $k$-SAT solutions  $\bm{s^*}\in{\cal V}_N$  are fixed points of  (\ref{1a}-\ref{1b}), {\it for any} $\bm{a}\in(0,\infty)^M$. The $k$-SAT solution {\it clusters} are spanning piecewise compact, connected sets in $Q_N$, and {\it every} point in them is a fixed point of (\ref{1a}-\ref{1b}) (Supplementary sect. A).
System (\ref{1a}-\ref{1b}) has a number of key properties (see Supplementary Information). (i) The dynamics in $\bm{s}$ stays confined to ${\cal H}_N$. (ii) The $k$-SAT solutions $\bm{s^*}\in{\cal V}_N$  are attractive fixed points of  (\ref{1a}-\ref{1b}). In particular, every point $\bm{s}$ from the orthant of a $k$-SAT solution $\bm{s^*}$ with the property $|\bm{s}|^2\geq N-1+(k-1)^2/(k+1)^2$  is guaranteed to flow into the attractor corresponding to $\bm{s^*}$.  (iii) There are no limit cycles.  (iv) For satisfiable formulae the {\it only} fixed point attractors of the dynamics are the global minima of $V$  with $V=0$. Note that in principle, the projection of the dynamics onto ${\cal H}_N$ could be stuck in some point $\overline{\bm{s}}$, while  $d\bm{a}/dt\neq0$ indefinitely. This does not happen here, as shown in Supplementary sect. E.  Moreover, analytical arguments supported by simulations indicate that the trajectory will leave any domain that does not contain solutions, see the discussion in Supplementary sect. E1.
Note, that the constraint functions (hence their satisfiability) depend directly only on the location of the trajectory in ${\cal H}_N$, $K_m=K_m(\bm{s})$, and not on the auxiliary variables. The dynamics in the $\bm{a}$-space is simple expansion, and for this reason the features of the full phase space $\Omega$ lie within its projection onto ${\cal H}_N$. One can actually eliminate entirely the auxiliary variables from the equations by first solving (\ref{1b}) to give 
$a_m(t)=a_m(0)\exp\left(\int_0^t K_m(\bm{s}(\tau))d\tau\right)$  then inserting it into (\ref{1a}).

 \begin{figure}[htbp]
\centerline{\includegraphics[width=5.30in]{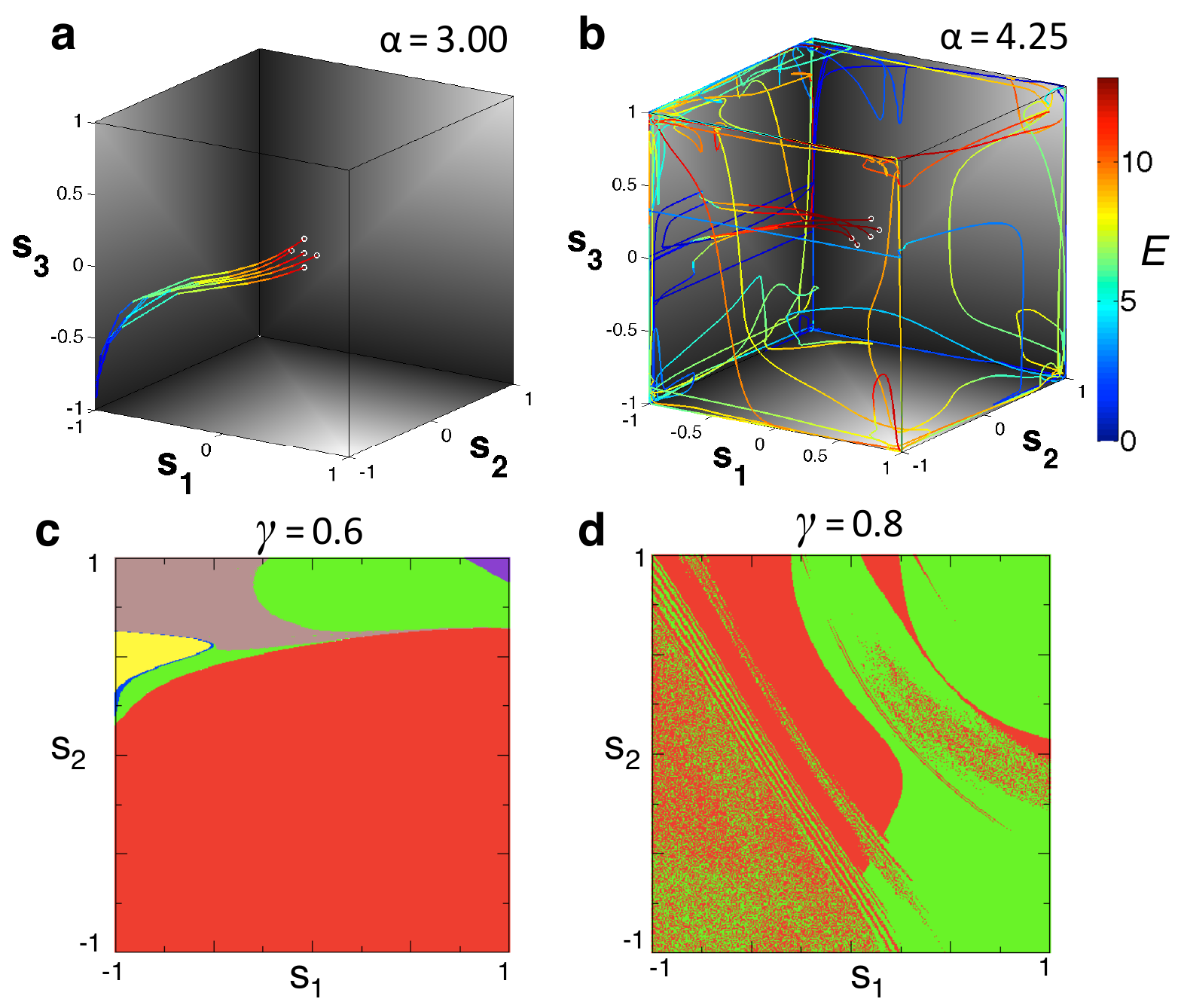}} 
\caption{{{\bf Chaotic behaviour}. Five, closely started sample trajectories projected onto  $(s_1,s_2,s_3)$ a), for a 3-SAT formula with $N=200$, $\alpha =3$ and b) for a hard formula, $N=200$, $\alpha=4.25$. The colour indicates the energy $E$ (colour bar) in a given point of the trajectory. While for easy formulae the trajectories exhibit laminar flow, for hard formulae they quickly become separated, showing a chaotic evolution. Taking a small $3$-XORSAT instance with $N =15$ (see Supplementary sect. G) we fix a random initial condition for all $s_i$, except $s_1$ and $s_2$ which are varied on a $400\times400$ grid and we colour each point according to  the solution they flow to for c) $\gamma =0.6$ (instance shown on Fig. 8e) and d) for $\gamma =0.8$ (instance shown in Fig. 8f)).}} \label{Fig1}
\end{figure}

\begin{figure}[htbp]
\centerline{\includegraphics[width=3.0in]{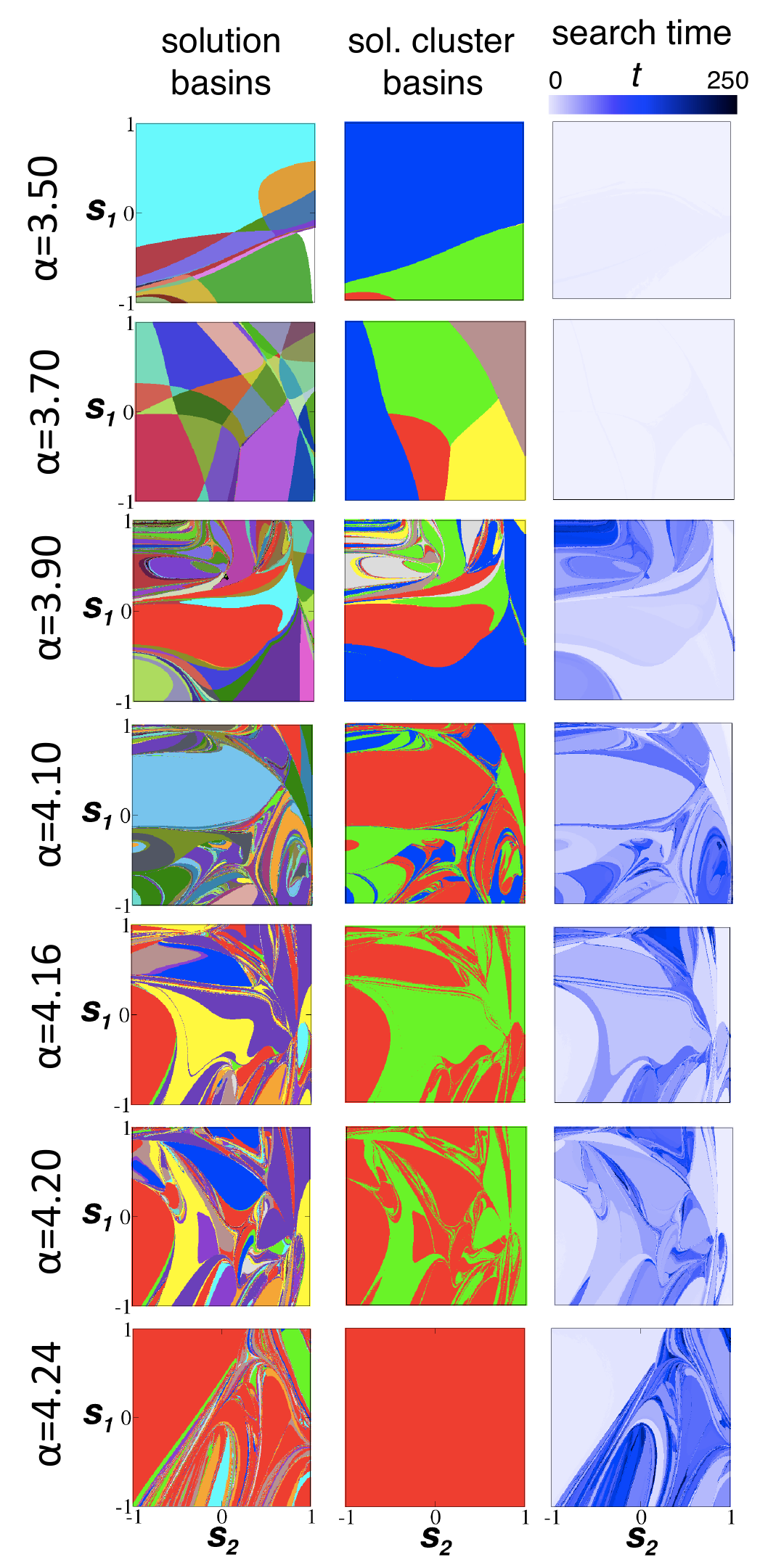}} 
\caption{{{\bf Attractor basins}. For a random $3$-SAT instance with $N=50$ we vary $\alpha$ by successively adding new constraints. Fixing a random initial condition for $s_i$, $i\geq 3$, we vary only $s_1$ and $s_2$ on a $400\times400$ grid, and we colour each point according to the solution (first column) or solution cluster (second column) they flow to.  Each colour in a given column represents a solution or solution cluster respectively, however colours between columns are independent. The third column represents the analog search time $t$ needed to find a solution (see colour bar) starting from the corresponding grid point. Maps are presented for values of $\alpha =3.5, 3.7, 3.9, 4.1, 4.16, 4.2, 4.24$. Easy formulae are characterized by smooth basin boundaries and small search times. Note that we only see the solutions (and clusters) that reveal themselves in the $(s_1,s_2)$ plane, others might not be seen. For hard formulae the boundaries and the search time maps become fractal.}} \label{Fig2}
\end{figure}

Another fundamental feature of (\ref{1a}-\ref{1b}) is that it is {\it deterministic}: for a given formula $f$, any initial condition generates a unique trajectory, and any set from ${\cal H}_N$  has a unique preimage arbitrarily back in time. Hence, the characteristics of the solution space are reflected in the properties of the invariant sets \cite{Ref7} of the dynamics (\ref{1a}-\ref{1b}) within the hypercube ${\cal H}_N$. The deterministic nature of (\ref{1a}-\ref{1b}) allows us to define basins of attractions of solution clusters by colouring every point in ${\cal H}_N$ according to which cluster the trajectory flows to, if started from there. These basins fill ${\cal H}_N$   up to a set of zero (Lebesgue) measure, which forms the basin boundary \cite{Ref7}, from where the dynamics (by definition) cannot flow to any of the attractors.  A $k$-SAT formula $f$ can be represented as a hypergraph
${\cal G}(f)$   (or equivalently, a factor graph) in which nodes are variables and hyperedges are clauses connecting the nodes/variables in the clause. Pure literals are those that participate in one or more clauses but always in the same form (direct or negated); hence they can always be chosen such as to satisfy those clauses. The core of ${\cal G}(f)$  is the subgraph left after sequentially removing all the hyperedges having pure literals \cite{Ref23}. For simple formulae (such as those without a core), the dynamics of (\ref{1a}-\ref{1b}) is laminar flow and the basin boundaries form smooth, non-fractal sets (Fig.\ref{Fig1}a,c and Fig.\ref{Fig2} top two rows). Adding more constraints  ${\cal G}(f)$  develops a core, the spin equations (\ref{1a}) become mutually coupled, and the trajectories may become chaotic (Fig.\ref{Fig1}b, Supplementary sect. F, Fig.12) and the basin boundaries fractal \cite{Ref7,Ref8,Ref9} (Fig. \ref{Fig1}d, Fig. \ref{Fig2}, Fig. 8). Therefore, as the constraint density $\alpha=M/N$  is increased within 
predefined ensembles of formulae (random $k$-SAT, occupation problems, $k$-XORSAT, etc.) a sharp change to chaotic behaviour is expected at a {\it chaotic transition point} $\alpha_{\chi}$, where a {\it chaotic} core appears with non-zero statistical weight in the ensemble as $N\rightarrow\infty$.  As an example, let us consider $3$-XORSAT. In this case, due to its inherently linear nature, it is actually better to work directly with the parity check equations as constraints, instead of their CNF form. The chaotic core here is a small finite hypergraph, and thus  $\alpha_{\chi}$ coincides with the so-called dynamical transition point  computed exactly by M\'ezard et al. \cite{Ref24}  (see Supplementary sect. G and Fig. 8). Note, a core can be non-chaotic, and thus the existence of a core is only a {\it necessary} condition for the appearance of chaos and in general the two transitions might not coincide. Further increasing the number of constraints (within any formula ensemble) unsatisfiability appears at the threshold value $\alpha_s>\alpha_{\chi}$  beyond which almost all formulae are unsatisfiable (UNSAT regime)\cite{Ref11,Ref12,Ref22,Ref24,Ref25,Ref26,Ref27,Ref28}.  The closer $\alpha$ is to 
$\alpha_s$, the harder it is to find solutions, and beyond the so-called freezing transition point   
$\alpha_f<\alpha_s$ (called the frozen regime) all known algorithms take exponentially long times or simply fail to find solutions \cite{Ref11,Ref12}. A variable is frozen if it takes on the same value for all solutions within a cluster, and a cluster is frozen if an extensive number of its variables are frozen. In the frozen regime all clusters are frozen and they are also far apart ( ${\cal O}(N)$ Hamming distance)\cite{Ref11,Ref12}. For random $3$-SAT (clauses chosen uniformly at random for fixed  $\alpha$) $\alpha_s\cong4.26$ \cite{Ref27}, $\alpha_f\cong4.25$ \cite{Ref28} and all known local search algorithms become exponential or fail beyond  $\alpha=4.21$ {\cite{Ref29}}, while Survey Propagation \cite{Ref25} based algorithms fail beyond  $\alpha=4.25$ \cite{Ref28}.  Since the frozen regime is very thin in random $3$-SAT, Zdeborov\'a and M\'ezard \cite{Ref12} have introduced the so-called locked occupation problems (LOPs).  In LOPs all clusters are formed by exactly one solution, hence they are completely frozen and the frozen regime extends from the clustering (dynamical) transition point  $l_d$ to the satisfiability threshold $l_s$, and thus it is very wide \cite{Ref12}. An example LOP is random Ò$+1$-in-$3$-SATÓ \cite{Ref12}, made of constraints that have no negated variables and a constraint is satisfied only if exactly one of its variables is $1$ (TRUE).  In $+1$-in-$3$-SAT  
$l_d\cong2.256$, $l_s\cong2.368$, and beyond $l_d$  all known algorithms have exponential search times or fail to find solutions (here $l=3M/N$). 
  \begin{figure}[htbp]
\centerline{\includegraphics[width=5.0in]{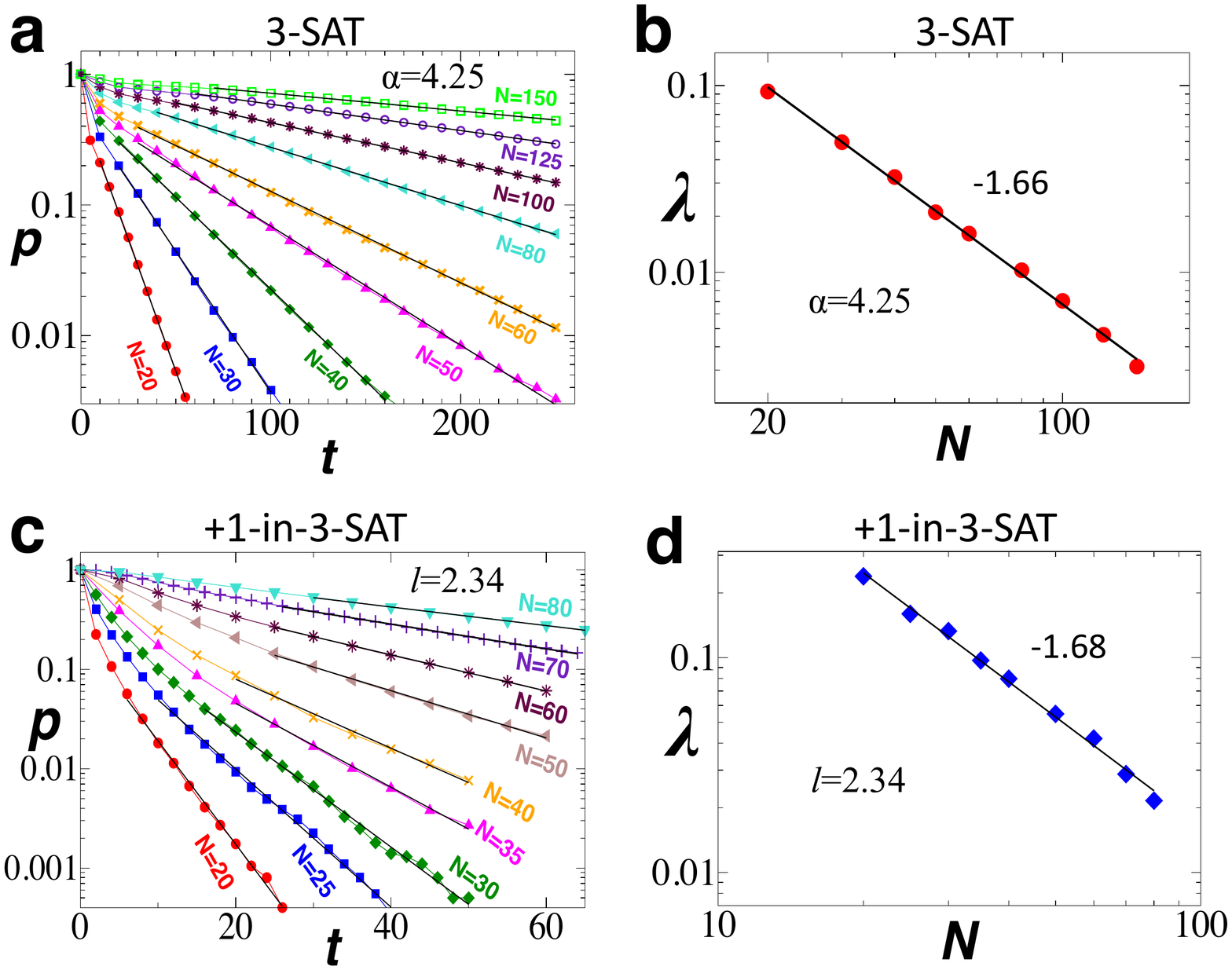}} 
\caption{{{\bf Computational complexity properties}. a) The fraction of problems $p(t)$ not yet solved by continuous-time $t$ for 3-SAT at $\alpha=4.25$, for $N=20,30,40,50,60,80,100,125,150$ (colours). Averages were done over $10^5$ instances for each $N$. For each instance the dynamics was started from one random initial condition. Black continuous lines show the decay $p(t)=r \exp(-\lambda(N)t)$. b) The decay rate follows  $\lambda(N)=bN^{-\beta}$, with $\beta\simeq 1.66.$ c) The fraction of problems $p(t)$ unsolved by time $t$ for $+1$-in-$3$-SAT at $l=2.34$, for $N=20,25,30, 35,40,50,60,70,80$. For each instance the dynamics was started in parallel from $10$ random initial conditions, averages were taken over $10^4$ instances for each $N$. Black continuous lines show the same exponential decay as in a). d) The decay rate shows the same behaviour as in b) with exponent: $\beta\simeq 1.68$. }} \label{Fig3}
\end{figure}    
 
As chaos is present for satisfiable formulae, that is, when system (\ref{1a}-\ref{1b}) \emph{has} attracting fixed points, it is necessarily of \emph{transient} type. Transient chaos \cite{Ref4,Ref5,Ref6,Ref7} is ubiquitous in systems with many degrees of freedom such as fluid turbulence \cite{Ref30}.  It appears as the result of homoclinic/heteroclinic intersections of the invariant manifolds of hyperbolic (unstable) fixed points of (\ref{1a}-\ref{1b}) lying within the basin boundary \cite{Ref7,Ref8,Ref9}, leading to complex (fractal) foliations of the phase space (see Supplementary sect. F). We observed the prevalence of transient chaos in the whole region 
 $\alpha_{\chi}<\alpha<\alpha_s$ for all the problem classes we studied. Interestingly, the velocity fluctuations of trajectories in the chaotic regime are qualitatively similar to those of fluid parcels in turbulent flows as shown in Supplementary sect. K.  Our findings suggest that chaotic behaviour may be a generic feature of algorithms searching for solutions in hard optimization problems, corroborating the observations by Elser et al.\cite{Ref10} using a heuristic algorithm based on iterated maps.

\begin{figure}[htbp]
\centerline{\includegraphics[width=6.0in]{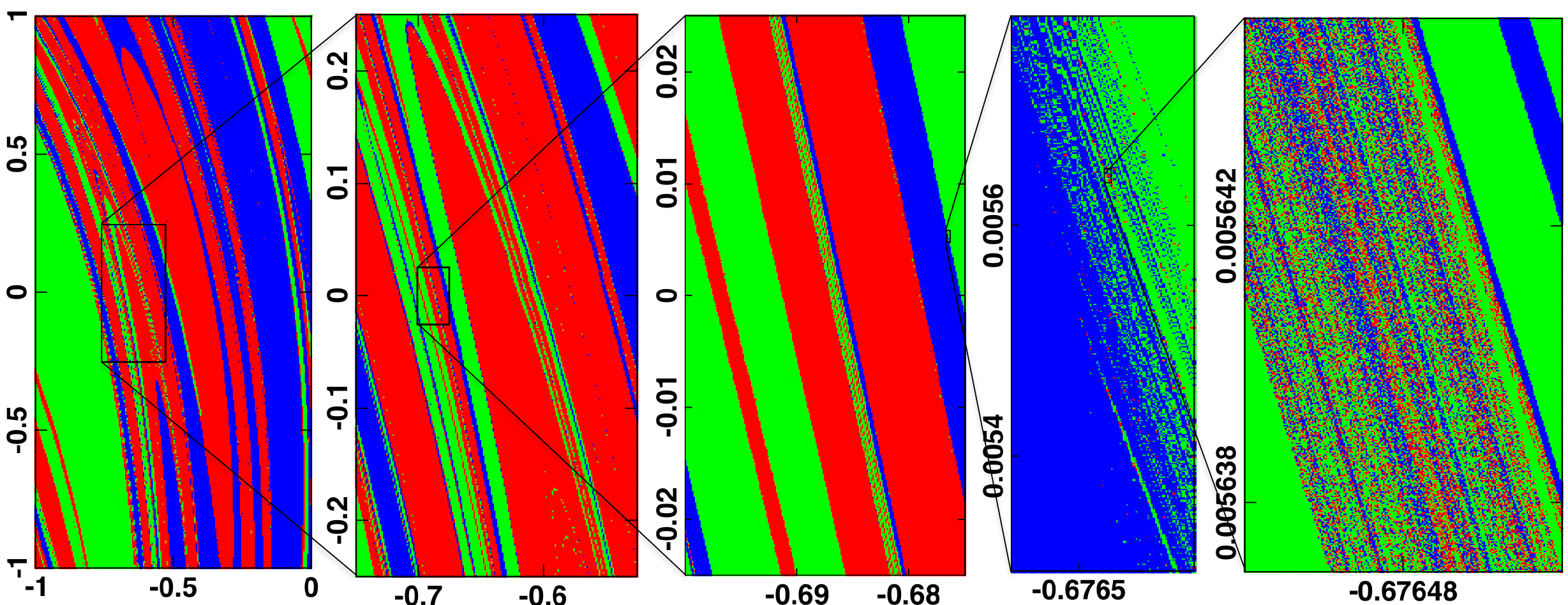}} 
\caption{{{\bf Wada property}.  Basin boundaries are shown for $+1$-in-$3$-SAT for an instance at $N=30$ and $l=2.28$. Fixing a random initial condition for all $s_i$, $i \geq 3$ we vary only $s_1$ and $s_2$ on a $200 \times 400$ grid and colour the points according to three different solutions they flow to. Successive magnifications illustrate the Wada property: the points on the basin boundaries are simultaneously on the boundary of all three basins implying that large enough magnifications will contain all three colours (although the blue-green boundary seems void of red in the third panel, panels four and five show that red is actually present).}} \label{Fig4}
\end{figure}

 In the following we show results on random $3$-SAT and $+1$-in-$3$-SAT formulae in the frozen regime, however, the same conclusions hold for other ensembles that we tested.  To investigate the complexity of computation by the flow (\ref{1a}-\ref{1b}), we monitored the fraction of problems  $p(t)$ not solved by continuous time $t$, as function of $N$ and $\alpha$. Figs. \ref{Fig3}a,c show that even in the frozen phase, the fraction of unsolved problems by time $t$ decays exponentially with $t$, that is, by a law $p(t)=r e^{-\lambda(N)t}$. The decay rate $\lambda(N)$ obeys $\lambda(N)=bN^{-\beta}$, with $\beta\simeq1.6$  in both cases, see Fig. \ref{Fig3}b,d. From these two equations, the continuous-time $t(p,N)$   needed to solve a fixed  $(1-p)$th fraction of random formulae (or to miss solving $p$-th fraction of them) is:  
  \begin{equation}
  t(p,N)=b^{-1}N^{\beta}\ln(r/p)  \label{tpn}
  \end{equation}
  indicating that the continuous time (CT) needed to find solutions scales as a power-law with $N$. Eq. (\ref{tpn}) also implies power-law scaling for almost all hard instances in the $N\rightarrow\infty$   limit (Supplementary sect. H). The \emph{length} in ${\cal H}_N$ of the corresponding continuous trajectories also scales as a power-law with $N$ (Supplementary Fig. 11b, sect. J). However, note that this does not mean that the algorithm itself is a polynomial-cost algorithm, as the energy function $V$ can have exponentially large fluctuations. As the numerical integration happens on a digital machine, it approximates the continuous trajectory with discrete points. Monitoring the fraction of formulae left unsolved as function of the number of discretization steps $n_{step}$ in the frozen phase, we find exponential behaviour for $n_{step}(p,N)$  (Supplementary sects. I,J, Fig. 10).  The difference between the continuous- and discrete-time complexities is due to the wildly fluctuating nature of the chaotic trajectories (see Fig. \ref{Fig1}b and Methods) in the frozen phase.  Compounding this, we also observe the appearance of the Wada property \cite{Ref7,Ref8} in the basin boundaries, Fig. \ref{Fig4}. A fractal basin boundary has Wada property if its points are simultaneously on the boundary of at least three colours/basins. (An amusing method that creates such sets uses four Christmas ball ornaments\cite{Ref7}.)  Although the Wada property does not affect the true/mathematical analog trajectories, owing to numerical errors, it may switch the numerical trajectories between the basins. Since the clusters are far (${\cal O}(N)$) apart, the switched trajectory will flow towards another cluster into a practically opposing region of ${\cal H}_N$ until it may come close again to the basin boundary etc., partially randomizing the trajectory in  ${\cal H}_N$.  
  
  We conjecture that the power-law scaling of the continuous search times (\ref{tpn}) is due in part to a generic property of the dynamical system (\ref{1a}-\ref{1b}), namely that it is \emph{hyperbolic} \cite{Ref4,Ref6,Ref13}, or near-hyperbolic. It has been shown that for hyperbolic systems the trajectories escape from regions far away from the attractors to the attractors at an \emph{exponential rate}, for almost all initial conditions \cite{Ref4,Ref6,Ref13}. That is, the fraction of trajectories still searching for a solution after time $t$ decays as   $e^{-\kappa t}$(Supplementary Fig. 13), where $\kappa$ is the \emph{escape rate}.  Thus, $\kappa^{-1}$  can be considered as a measure of hardness for a \emph{given formula}. When taken over an ensemble at a given $\alpha$, this property generates the exponential decay for $p(t)$  with an average escape rate $\lambda$.  
  
 The form of the energy function $V$ incorporates the influence of all the clauses at all times, and in this sense the system (\ref{1a}-\ref{1b}) is a non-local search algorithm.  As shown before, the auxiliary variables can be eliminated, however, they give a convenient interpretation of the dynamics. Namely, one can think of them as providing extra dimensions along which the trajectories escape from local wells, and their form (\ref{1b}) provides positive feedback that guarantees their escape. Clearly, these equations are not unique, and other forms based on the same principles may work just as well.

\subsection*{Methods}
  
To simulate (\ref{1a}-\ref{1b}), we use a 5-th order adaptive Cash-Karp Runge-Kutta method with monitoring of local truncation error to ensure accuracy. In order to keep the numerical trajectory within a tube of small, preset thickness around the true analog trajectory in $\Omega$   (Supplementary Fig. 9), the RK algorithm occasionally performs an exponentially large number of discretization steps $n_{step}$. However, this only happens for hard formulae, when the analog trajectory has wild, chaotic fluctuations. For easy formulae both $p(t)$ and $p(n_{step})$  decay exponentially as shown in Supplementary Fig. 10a, inset.  
  
\subsection*{Acknowledgments} 
We thank T. T\'el and L. Lov\'asz for valuable discussions and for a critical reading of the manuscript.  

\subsection*{Author contributions}
M.E.R. and Z.T. conceived and designed the research and contributed analysis tools equally. M.E.R. performed all simulations, collected and analysed all the data and Z.T. wrote the paper.


\appendix
\section{Supplementary Information}
\nid  Here we provide derivations and discussions about the properties of the 
dynamical system (\ref{1a}-\ref{1b}) and additional supporting figures and text. 

\medskip

\noindent Recall from the main text the definitions:
\begin{equation}
V(\bm{a},\bm{s})= \sum_{m=1}^{M} a_m K_m^2\;, \label{V}
\end{equation}
where
\begin{equation}
K_m= 2^{-k}\prod_{j=1}^{N} \left(1- c_{mj}s_j\right) \label{Km}\;.
\end{equation}
Note that in $k$-SAT there are at most $k$ terms in the product above, 
hence $0 \leq K_m \leq 1$
for all $m$.
The system of ODEs (\ref{1a}-\ref{1b}) defined in the main document is:
\begin{eqnarray}
&&\dot{s}_i = \frac{ds_i}{dt} = - \frac{\partial}{\partial s_i} V(\bm{a},\bm{s})= \sum_{m=1}^{M} 
2 a_m c_{mi} K_{mi} K_m\;,\;\;\;\;i=1,\ldots,N\;, \label{sdyn}  \\
&&\dot{a}_m = \frac{da_m}{dt} =  a_m K_m\;,\;\;\;\;m=1,\ldots,M\;, \label{adyn}
\end{eqnarray}
with $a_m(0) > 0$, $m=1,\ldots,M$ ( for example $a_m(0) = 1$), and where
\bq
K_{mi} = 
2^{-k}\prod_{\stackrel{j=1}{j\neq i}}^{N} \left( 1-c_{mj} s_j \right) = 
\frac{K_m}{1-c_{mi}s_i}\;.\label{kmi}
\eq

\begin{figure}[htbp]
\centerline{\includegraphics[width=5.30in]{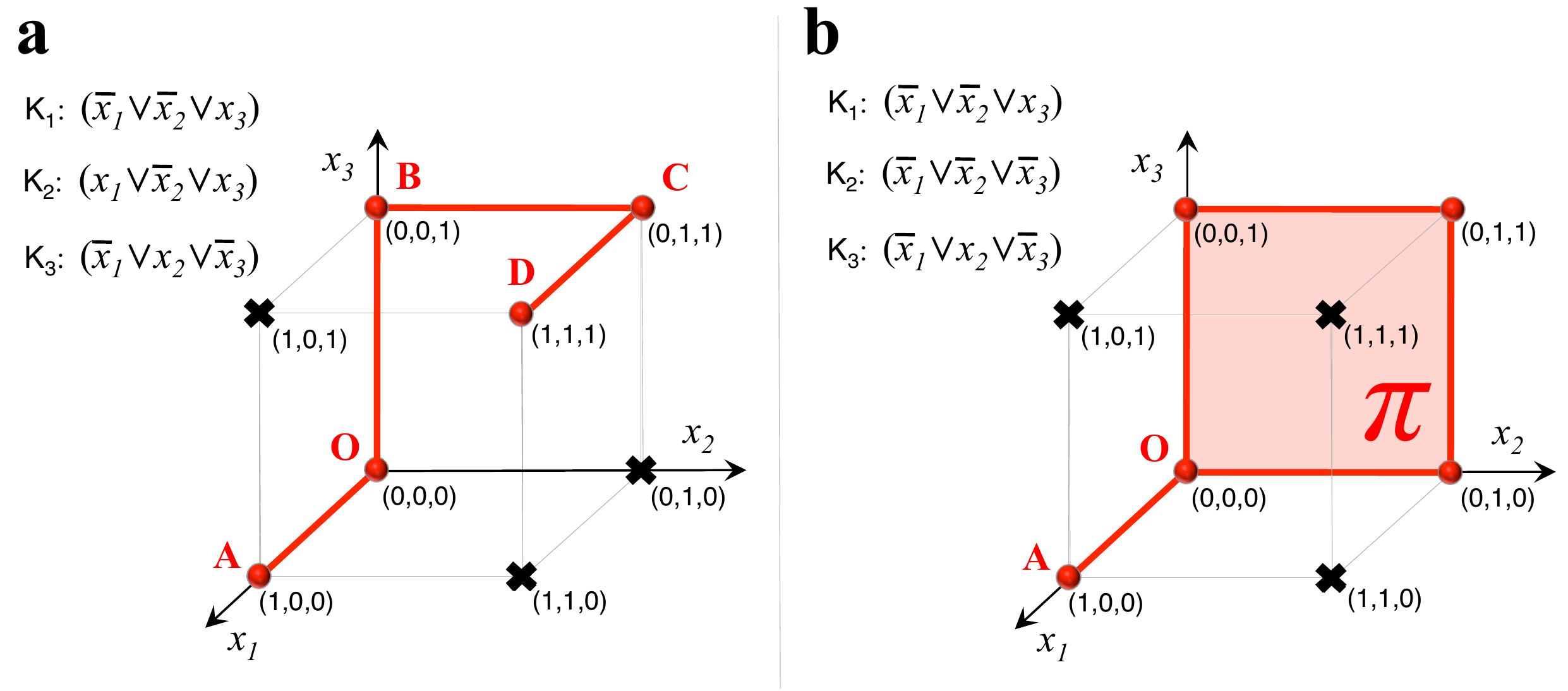}} 
\caption{{\bf Example attractors spanned by solution clusters.} 
Two examples with $k=3$, $N=3$, $M=3$, see the figure for the 
sets of clauses in each case. In {\bf a} the solution cluster is the set 
of vertices $\{A,O,B,C,D\}$, while for the dynamical
system the corresponding attractor is the {\em continuous} union of segments 
$\overline{AO}\cup\overline{OB}\cup\overline{BC}\cup
\overline{CD}$ (and the equivalent in the $\bm{s}$-space). 
In {\bf b}, the solution cluster is 
formed by the four vertices of the shaded square-plate $\pi$ and
vertex $A$, while for the dynamical system the corresponding 
attractor is $\pi \cup \overline{AO}$. } \label{fig:S1}
\end{figure}

\paragraph{A. Free variables, solution clusters and attractors.} Solution clusters are 
defined by solutions that can be connected via single-variable flips, always staying 
within satisfying assignments. For example, consider two k-SAT solutions that differ 
in exactly one variable, let's say in $s_j$ (at Hamming distance of 1), thus forming 
a solution cluster of two points in ${\cal V}_N = \{-1,1\}^{N}$. Then {\em any} point 
on the $s_j$ axis in the $[-1,1]$ 
{\em continuous segment} is a fixed-point ($\dot{\bm{s}}=\bm{0}$, $\dot{\bm{a}}=\bm{0}$) 
of the dynamics (\ref{sdyn}-\ref{adyn}), because for the two solutions the value 
of $s_j$ is irrelevant 
(called a ``free variable''), all the clauses being satisfied by the other variables taking 
values  $\pm1$. For our dynamical system, the corresponding attractor is an edge
of the $N$-hypercube $Q_N$, a compact, connected domain/set.  It is certainly possible 
that several solutions in ${\cal V}_N$  be connected via single variable flips, with 
two examples given in Figs. \ref{fig:S1}a-b. Thus, the {\em attractors} of the dynamical 
system  (\ref{sdyn}-\ref{adyn}) (or (1) in the main text) in general are compact, 
connected sets of  ${Q}_N$ spanned by the k-SAT solution clusters  and 
by only those (see also sections D and E). Every point from this attractor set is a 
fixed-point of the continuous-time (CT) dynamics. 

\paragraph{B. Spin variables $\bm{s}$ remain in  ${\cal H}_N=[-1,1]^N$.}
From Eq. (\ref{sdyn}):
\begin{eqnarray}
 \dot{s}_i=  2\sum_{m=1}^{M} a_m c_{mi} (1-c_{mi} s_i) K_{mi}^2 = 2 
\sum_{m=1}^{M} a_m K_{mi}^2 \left[(1-s_i)\delta_{c_{mi},1} - 
(1+s_i)\delta_{c_{mi},-1} \right] \quad \label{sdot}
\end{eqnarray}
This dynamics keeps spins within $[-1,1]$, because at $ s_i =1$ ($s_i = -1$) we have 
$\dot{s}_i \leq 0$ ($\dot{s}_i \geq 0$) for any $c_{mi} \in \{-1,1\}$.   

\paragraph{C. Stability of all $k$-SAT solutions. Domains of attraction.}
As seen in the main text, all the $\bm{s}$ points for which $V=0$, are fixed 
points of the dynamics (\ref{sdyn}-\ref{adyn}). 
As shown in Section A, some of these points are not necessarily 
from ${\cal V}_N$, they can be from the compact domain corresponding to 
a cluster of solutions from ${\cal V}_N$, when there are free variables.   
We prove stability by showing that in a vicinity of non-zero volume 
in ${\cal H}_N$ of a $k$-SAT solution,  the function 
$R\equiv\sum_{i=1}^N {s_i^2}$
is monotonically increasing until the dynamics reaches a point in 
$\bm{s}$-space for which  $V=0$ (hence all $K_m = 0$), that is, the trajectory
reaches the cluster's domain. 
From (\ref{sdyn}):
\bq
\dot{R} \equiv 2\sum_{i=1}^{N}  s_i \dot{s}_i=4\sum_{m=1}^{M}{a_m K_m 
\sum_{i=1}^{N}{c_{mi} s_i K_{mi}}} \,\, . \label{dRperdt}
\eq
Assume that $\bm{s^*}\in {\cal V}_N$ is a $k$-SAT solution. Recall that we have $k$ variables  
in each clause (all the rest have $c_{mi} = 0$).  
Choosing a clause $K_m$, let us denote the indices of the variables included as $i=1,\dots, k$,
and order them such that the variables  $j=1,\dots,p$ satisfy the clause, meaning 
that $(1-c_{mj}s^*_j)=0$,
and the  variables $l=p+1,\dots,k$ do not satisfy the clause, 
and thus $(1-c_{ml}s_l^*)=2$. Clearly, such a $1 \leq p \leq k$ always exists, since 
$K_m(\bm{s^*}) = 0$. If we are in the corresponding $N$-dimensional octant (orthant) 
of $\bm{s^*}$ then 
$sgn\{s_i\}=s_i^*$ for all $\bm{s}$ from this orthant with $0 < |s_i | < 1$. 
Then $c_{mj}s^*_j=1$ implies  $c_{mj}s_j=|s_j|$ for all $j=1,\dots,p$ and $c_{ml}s_l^* = -1$ 
implies $c_{ml}s_l=-|s_l|$ for all $l=p+1,\dots,k$ in this orthant.  Thus:
\bq
Z_m \equiv \sum_{i=1}^k{c_{mi} s_i K_{mi}} = K_m \sum_{i=1}^k{\frac{c_{mi}s_i}{1-c_{mi}s_{i}}}= 
K_m \left(\sum_{j=1}^p{\frac{|s_j|}{1-|s_j|}}-\sum_{l=p+1}^k{\frac{|s_l|}{1+|s_l|}}\right) \label{ineq1}
 \eq
Clearly, for any $\sigma > 0$ with $\sigma < \min_{j} |s_j|$, we have $|s_j|/(1-|s_j|) > 
\sigma/(1-\sigma)$ (for all $j$) and $|s_l|/(1+|s_l|) < 1/2$ (since $|s_l| < 1$). Therefore
\bq
Z_m > K_m \left( \frac{p \sigma}{1-\sigma} - \frac{k-p}{2}\right)\;. \label{ziem}
\eq
\begin{figure}[htbp] 
\centerline{\includegraphics[width=2.50in]{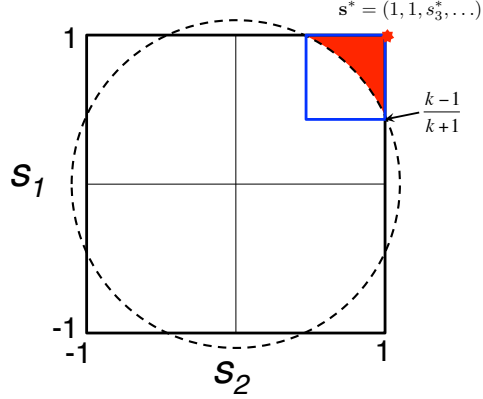}} 
\caption{{\bf Domain of attraction, 2D illustration.} If $\bm{s^*}\in \cal{V}_N$ 
is a k-SAT solution, and  $sgn(s_i)=s_i^*$, 
$|s_i|>(k-1)/(k+1)$ are satisfied for all $i=1,\dots,N$ (inside 
the blue square) then $dR/dt>0$. The dynamics cannot leave from inside 
the domain ${\cal C}_N(\bm{s^*})$ (marked with red) as $|s_i| \leq 1$ 
(see section B). It follows that $R$ grows 
until it hits the domain of the solution cluster attractor containing $\bm{s^*}$, or $\bm{s^*}$ itself, 
meaning that  ${\cal C}_N(\bm{s^*})$ is part of the attractor's basin.} \label{fS1}
\end{figure}
Hence,  if $\sigma > \frac{k-p}{k+p}$ , then $Z_m > 0$ (for all $m$) 
and from
(\ref{dRperdt}) $\dot{R}   = \sum_{m} 4a_m K_m Z_m > 0$, that is the trajectory is 
strictly increasing its distance from the origin
(unless all $K_m = 0$, which means the trajectory is on the attractor, where $\dot{R}=0$).
Since $(k-p)/(k+p)$ is a decreasing function of $p$, we can choose $p = 1$ and set
$$
\sigma = \frac{k-1}{k+1}
$$
to define a vicinity of $\bm{s^*}$, with $|s_i | > \sigma$, $i=1,\ldots,N$ 
within which we are guaranteed $\dot{R} > 0$. 
Next, consider  
the corner domain  ${\cal C}_N(\bm{s^*})$ around $\bm{s^*}$ of ${\cal H}_N$ 
{\em cut away} by the $N$-dimensional sphere of radius $\sqrt{N-1 + \sigma^2}$
centered in the origin. See Fig. \ref{fS1} for an illustration in 2D. 
Trajectories in all points within this domain have the property 
$\dot{R} > 0$ or $\dot{R} = 0$, in the latter case the point being on the attractor. For points with 
$\dot{R} > 0$, the trajectory must necessarily flow towards the boundary part ($Q_n$)
of the hypercube of this domain ${\cal C}_N(\bm{s^*})$,
away from the surface of the sphere until it hits  $\bm{s^*}$, or the attractor of the solution
cluster of which $\bm{s}^*$ is part of (in case of free variables), lying within ${\cal C}_N(\bm{s^*})$. 
Clearly $R$ cannot
increase beyond $N$, in which case the trajectory is at $\bm{s^*}$.

\paragraph{D. There are no limit cycles in $\bm{s}$.}
Having a limit cycle in $\bm{s}$ which is not a fixed-point means that 
$s_i(t)$ is a periodic function of $t$ of period $T > 0$, 
and consequently all its derivates, including $\dot{s}_i(t)$:
\bq
s_i(t) = s_i(t+n T)\;,\;\;\;\;\dot{s}_i(t) = \dot{s}_i(t+n T) \label{dscycle}
\eq
for all $i=1,\dots,N$ and all integers $n$. Since the $K_m$ are functions 
of time only through the $s_i$ variables, this implies
that all $K_m$ are periodic functions of time as well. From (\ref{sdyn}) it follows:
\bq
\dot{s}_i(t+n T) =\sum_{m=1}^{M} a_m(t+n T) 2 c_{mi} 
K_m(\bm{s}(t)) K_{mi}(\bm{s}(t)).   \label{dstplusbetaT}
\eq
Formally, the solution for $a_m$ from (\ref{adyn}) can be written as
\begin{equation}
a_m(t) = a_m(t_0) \exp \left[ {\int_{t_0}^{t} d\tau K_m(\bm{s}(\tau))} \right]. \label{asol}
\end{equation}
Using the periodicity of $K_m$ this leads to
$a_m(t+n T)=a_m(t) e^{n I_m}$
with 
$I_m=  \int_{0}^{T}{ K_m(\bm{s}(\tau)) d\tau}$.
From (\ref{dstplusbetaT}) and (\ref{dscycle}) it follows:
\bq
\sum_{m=1}^{M} a_m(t) \left(e^{ n I_m}-1 \right) 2 c_{mi} K_m(t) K_{mi}(t)=0\;,\;\;\;
i=1,\dots,N  \label{condition}
\eq
which has to hold for any integer $n$ and all times $t$. Every sum in (\ref{condition})
is of the type
\bq
\sum_{m} \left( x_m^n -1 \right) f_{mi} = 0\;, \;\;\;\;\forall\;\;n\in \mathbb{Z}\;, \label{type}
\eq
Where $x_m = e^{I_m}$. Since $K_m \in [0,1]$ it follows that $I_m \geq 0$ and thus $x_m \geq 1$.
Assume that the $x_m$-s are all different from each other, and consider $x_{m^*}$ to be the largest of them. Then, 
clearly, increasing $n$ without limit, all other terms in the sum of (\ref{type}) become arbitrarily small
compared to the $m^*$ term and since $x_{m^*} \geq 1$, this forces $f_{m^*i} = 0$. Thus the $m^*$
term must be absent from the sum (\ref{type}), and we can repeat the procedure with the next largest
$x_m$ term leading to $f_{mi} = 0$, etc. Hence, all $f_{mi} = 0$ and thus 
$\dot{s}_i = \sum_m f_{mi} = 0$. If the $x_m$-s are equal ``in blocks", with a similar procedure we
can show that for  every block $b$ we must have  $\left( x_b^n -1\right)\sum_{m\in b} f_{mi} = 0$, 
for all $n$. This implies that either $x_b = 1$, or $\sum_{m \in b} f_{mi} = 0$. In the former case we thus
must have for all $m \in b$, $I_m = 0$ and hence $K_m = 0$ (recall, that $K_m \geq 0$) which actually
implies that the corresponding $f_{mi} = 0$, since $f_{mi} \propto K_m$. Hence, either way, $\sum_{m\in b} f_{mi} = 0$ for all blocks, and therefore, again, $\dot{s}_i = \sum_m f_{mi} = 0$
meaning that we are in a fixed point, contradicting our original assumption.

\paragraph{E. Fixed-points and attractors in ${\cal H}_N$.} We have seen that the dynamics 
in $\Omega$ admits fixed point solutions ($\dot{\bm{s}} = \bm{0}$, $\dot{\bm{a}} = \bm{0}$) 
only on the boundary $Q_N$, either as isolated points from ${\cal V}_N$ ($k$-SAT solutions), or 
in form of compact, connected domains (if there are free variables) corresponding to $k$-SAT
solution {\em clusters}. This holds because $\dot{\bm{a}} = \bm{0}$ if and only if 
$K_m = 0$ for all $m$, which 
is possible (due to their product form, (\ref{Km})) only if some of the spins are $\pm 1$,  
hence they form $k$-SAT solutions.
The question is whether there are other {\em stable} fixed points {\em within} ${\cal H}_N$ (thus in 
$\bm{s}$-space), but with not all $K_m = 0$, in which the dynamics could get stuck indefinitely. 
In principle it could happen that there are points $\bm{s} \in {\cal H}_N$ for which 
$\dot{\bm{s}} = \bm{0}$ but the dynamics (\ref{adyn}) of the auxiliary variables is not able to 
unstuck the spin variables from there. The answer is negative that is, there are no 
such {\em stable} fixed points and here we sketch its proof. 
However, as we will see, there can be unstable fixed-points, 
which play an important role in the dynamics within ${\cal H}_N$.
In subsection E.1 
below we discuss how the trajectory must leave any domain ${\cal D}$ within ${\cal H}_N$
that does not have a solution in it.

Let us assume that the dynamics at time $\bar{t}$ arrives into a point 
$\bm{\bar{s}} = \bm{s}(\bar{t})\in {\cal H}_N$ for which:
\bq
\dot{s}_i(\bar{t}) = - \frac{\partial}{\partial s_i} V(\bm{a},\bm{s}) \Bigg|_ {\bar{s}_i} = 
2\sum_{m=1}^{M} a_m(\bar{t}) c_{mi} K_m(\bm{\bar{s}}) K_{mi}(\bm{\bar{s}}) = 
0\;,\;\;\;\;i=1,\ldots,N\;, \label{scond}
\eq
and not all $K_m$-s are zero.
Let us denote $\bar{a}_m \equiv  a_m(\bar{t})$, $\bar{K}_m \equiv K_m(\bm{\bar{s}})$. Then,
using (\ref{kmi}), system (\ref{scond}) can be thought of as a homogeneous system of equations for 
the $\bar{a}_m$ variables:
\bq
 \sum_{m=1}^{M} \bar{a}_m \bar{K}^2_m 
\frac{c_{mi}}{1-c_{mi}\bar{s}_i} = \sum_{m=1}^{M} \bar{a}_m \bar{u}_{mi}  = 
0\;,\;\;\;\;i=1,\ldots,N\;. \label{hcond}
\eq
As there are more clauses than variables ($M > N$), system (\ref{hcond}) in general
defines an $M-r$-dimensional domain in $\bm{a}$-space, namely, the {\em left null-space}
$\ker\!\left(\bm{U}^{\sf T} \right)$ of the $M\times N$ matrix $\bm{U} = \{u_{mi} \}$ of rank $r \leq N$. 
Since not all $K_m$-s are 
zero, the $a_m$ variables will change over time, according to (\ref{adyn}). In order for the 
dynamics to be stuck in $\bm{\bar{s}}$ indefinitely, one must have condition (\ref{hcond})
hold for {\em all later times} $\bar{t}+ \tau$. Using $a_{m}(t) = \bar{a}_m 
e^{\int_{\bar{t}}^{\bar{t}+\tau} dt \bar{K}_m} = \bar{a}_m e^{\tau \bar{K}_m}$, the condition becomes:
\bq 
\sum_{m=1}^{M} e^{\tau \bar{K}_m} \bar{a}_m  \bar{u}_{mi} =
0\;,\;\;\;\;i=1,\ldots,N\;,\;\;\;\mbox{for all}\;\;\;\tau \geq 0\;. \label{hcondt}
\eq
If there is a {\em single} $m^*$ for which $\bar{K}_{m^*} > \bar{K}_m $ for all $m$, 
then clearly, the lhs of 
(\ref{hcondt}) becomes dominated by this maximum exponential term and (\ref{hcondt}) 
becomes violated, and hence the dynamics becomes unstuck from $\bm{s}$. 
The only way that (\ref{hcondt}) has a chance to hold for arbitrary 
$\tau$, if the $K_m$-s are equal in blocks
of at least size two (since $ \bar{u}_{mi} \propto c_{mi} \in \{-1,0,1\}$, cancellations are possible).  That means that there are at least $M/2$ equations expressing the equalities
of the corresponding $K_m$-s. Since there are $N$ spin variables and we are looking at 
$\alpha = M/N > 2$ (no point solving SAT in the very easy phase) there are more (nonlinear) 
equations than variables and thus it
 drastically reduces the
chances of finding an $\bm{\bar{s}}$ solution satisfying these. Assuming that such 
$\bm{\bar{s}}$ exists (one cannot exclude it in general), we need
to analyze this case, in particular whether the equality in blocks of the $K_m$-s is an 
``attractive'' condition by the dynamics for some $\bm{\bar{s}}$. In other words, we
need a stability analysis of $\bm{\bar{s}}$. It is important to note that $\bm{\bar{s}}$ 
is not a standard fixed point in the traditional sense. Since the dynamics is also driven by the
auxiliary variables, we have the continuous domain of the left null-space 
$\ker\!\left(\bm{U}^{\sf T} \right)$ as ``fixed point'' 
for the dynamics. As soon as the $a_m$ variables leave this domain, the 
dynamics gets unstuck
from $\bm{\bar{s}}$, and only then. For this reason, the linear stability analysis around 
$\bm{\bar{s}}$ is somewhat different from a standard stability analysis. 

Thus, let us assume that (\ref{hcondt}) holds for arbitrary $\tau$ for some $\bm{\bar{s}}$. 
Since the dynamics is in the $(\bm{s},\bm{a})$-space, let us consider a small deviation 
$\bm{\epsilon}$ such that  $\bm{a} = \bm{\bar{a}} +\bm{\epsilon}$,  $\epsilon_m \geq 0$, 
$m=1,\ldots,M$,  $|\bm{\epsilon}| \ll 1$, at $\tau = 0$. 
Clearly, one can easily choose the $\epsilon_m$ variables, such that
\bq 
\sum_{m=1}^{M} \epsilon_m  \bar{u}_{mi} \neq 0
\;,\;\;\;\;i=1,\ldots,N\;, \label{ec1}
\eq
that is, $\bm{\epsilon}$ is any vector that is not from the left null-space of the 
$\{ \bar{u}_{mi}\}$ matrix. For example, if $\bar{u}_{11} \neq 0$, then 
$\bm{\epsilon} = (\varepsilon|\bar{u}_{11}|,0,\ldots,0)$, $\varepsilon >0$, $\varepsilon \ll 1$ would suffice. 
From (\ref{scond}) it then follows, that the $s_i$ variables start
changing, as $\dot{s}_i = 2 \sum_m \epsilon_m \bar{u}_{mi} \neq 0$. Recall that in $k$-SAT
we are only considering problems in which every variable is present in both its direct and 
negated forms, in different clauses. 
If this wasn't the case, that is, if a variable $s_i$ 
when present in a clause would be only present in its direct form  (no negated form 
anywhere), then we could easily set the value
of that variable to +1 and automatically satisfy all the clauses in which it is present and reduce the
problem to a smaller one. Additionally, if all $c_{mi}$-s would have the same sign, one could never have (\ref{hcond})
satisfied in the first place, since $\bar{a}_m > 0$ and $1-c_{mi} \bar{s}_ i > 0$. This means,
that for any $\dot{s}_i \neq 0$, there will be clauses for
which the $K_m$-s will increase and others for which it will decrease. Since these variations are continuous, the $\bm{\epsilon}$ shift vector can always be chosen such that 
there will appear a single largest clause $K_{m^*}$, resulting in the case already discussed above, 
with  $\bm{a}$ leaving exponentially fast the left null-space $\ker \!\left(\bm{U}^{\sf T} \right)$ 
showing that the dynamics is unstable in $\bm{\bar{s}}$. 

Clearly, it is in general mathematically possible for the $K_m$-s to be 
equal in blocks such that (\ref{hcondt}) holds for all $\tau$, 
defining regions of zero (Lebesgue) measure in ${\cal H}_N$. As we have
just shown, these fixed regions or ``points'', however, are all unstable for the dynamics. The linearized
neighborhoods of these points are characterized by the stable and unstable subspaces
spanned by the eigenvectors with contracting and expanding eigenvalues, respectively \cite{Ott00}.
For this reason, these type of fixed points are called hyperbolic fixed points or saddles. 
Moving away from the linearized neighborhoods, these two subspaces form the stable and unstable
manifolds of the saddle, which then extend endlessly into the phase space. While two stable and two
unstable manifolds can never cross, the stable and unstable manifolds can intersect, forming
either homoclinic intersections (when the two manifolds belong to the same saddle) or heteroclinic
intersections (they come from two different saddles), see Ref \cite{Ott00}. Chaotic dynamics
appears as the result of homoclinic or heteroclinic intersections.

\ms

\nid {\bf E.1 Escape from an arbitrary domain.} So far  we have shown an 
important property, that is, the dynamics cannot be captured by 
a non-solution {\em fixed point} (for any formula). 
We have also shown in Section D that it cannot be captured by a limit cycle either. 
The question remains, however, whether it could be captured by some other type of 
non-solution attractor, 
possibly even a chaotic attractor. 
One way to show that this cannot happen, is to prove that the trajectory cannot stay confined for arbitrarily long times within an {\em arbitrary 
domain} ${\cal D}$ in the $\bm{s}$-space (${\cal D} \subset {\cal H}_N$) that {\em does not 
contain a solution}. The following description is a brief sketch  for what
happens during the dynamics, and it forms the elements of a proof 
(to be published elsewhere). 

Let us assume that at some point in time $t_0$, the trajectory is in 
${\cal D}$, $\bm{s}(t_0)\in {\cal D}$.  After eliminating the auxiliary variables from (\ref{sdyn})
using their expression from (\ref{asol}) ,
the rate of change for the spin variables becomes: 
\begin{equation}
 \frac{ds_i}{dt} = \sum_{m=1}^{M} 
2 a_m(t_0)c_{mi} K_{mi} K_m \; e^{\int_{t_0}^{t} d\tau K_m(\bm{s}(\tau))}\;. \label{sdyne} 
\end{equation}
As $\bm{s}$ is confined to ${\cal D}$, there is always a subset of 
constraints ($K_m$-s) that are bounded away from zero within ${\cal D}$ 
(otherwise we are in a solution,
which contradicts the assumption that ${\cal D}$ has no solutions in it). After long enough times, the
exponentials in (\ref{sdyne}) involving these constraints grow very large, and unless they
balance each other perfectly (due to the $c_{mi}$-s) {\em in every} $\bm{s}$ point along the 
trajectory in ${\cal D}$, one of them will overtake the others and overgrow them 
($\exp \left[ \int_{t_0}^{t} d\tau K_m \right] > e^{K (t-t_0)}$), at least for a while. 
In this time interval the sign for some of the $ds_i/dt$-s
will stabilize into either just positive or just negative and their magnitude will grow
{\em exponentially fast} ($\left| Ae^{at}-Be^{bt}\right| \sim e^{\max\{a,b\}t}$). 
When that happens, at large enough $t$, the trajectory will outburst
from ${\cal D}$ (a finite domain). It can also happen, however, that the constraints oscillate 
around each other, and hence their role as to which one is the leading one in (\ref{sdyne}), 
changes over time. This can result in an oscillatory behavior in
the corresponding $ds_i/dt$-s and of the trajectory. However, even in this case, all 
terms corresponding to the constraints are growing exponentially fast, and the 
differences between the growing exponentials have increasingly wilder fluctuations, which
eventually leads to the outburst of the trajectory from ${\cal D}$.  
One can show that any situation in which the exponential terms corresponding to 
the constraints in all the $N$ equations perfectly cancel each
other (so that there is no leading exponential) 
will be unstable against small perturbations, similarly to the case of a fixed point.  

\begin{figure}[htbp]
\centerline{\includegraphics[width=6.2in]{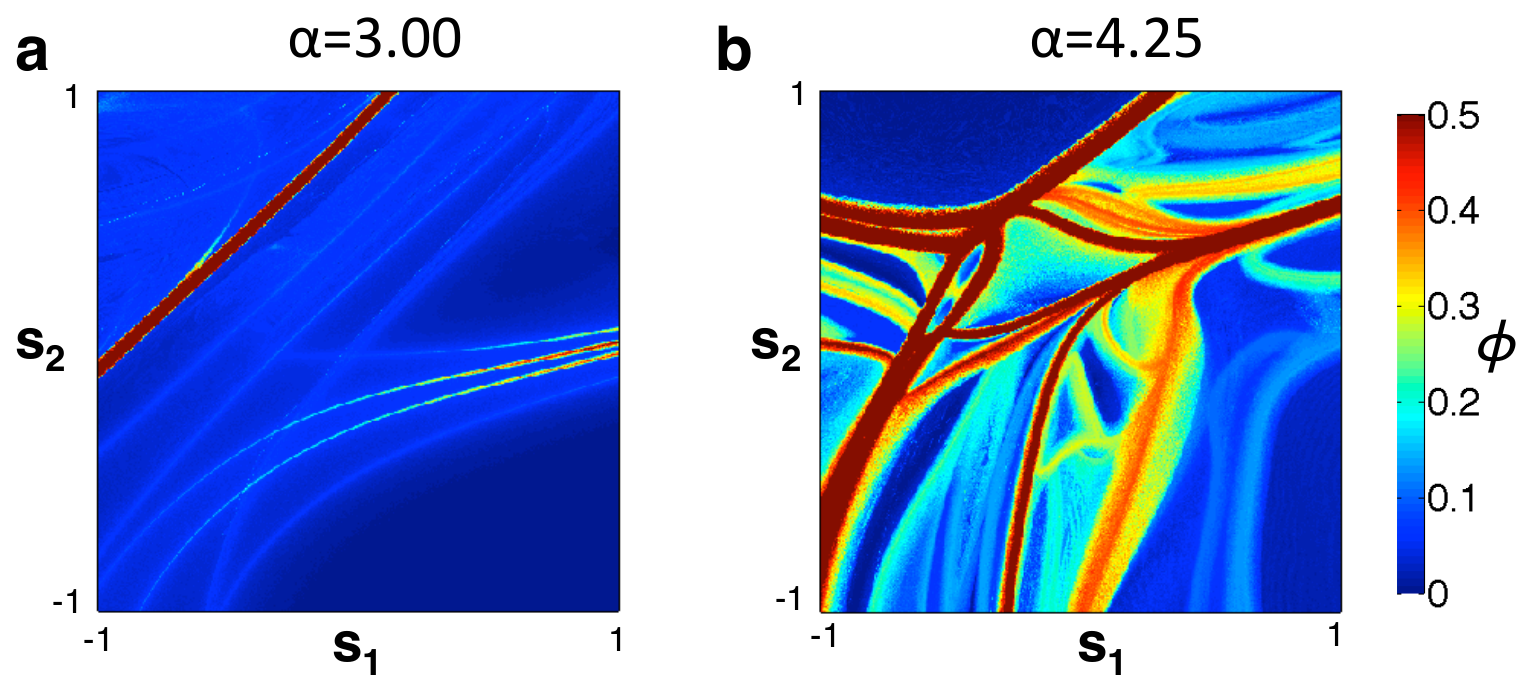}} 
\caption{{\bf Finite-Size Lyapunov Exponents.} 
The colour maps show Finite Size Lyapunov  Exponents $\phi$ measured on a problem with $k=3$, $N=50$ at {\bf a:} $\alpha=3$ and {\bf b:} $\alpha=4.25$. 
Initially, all $s_j$ have a random, but fixed value except $s_1$ and $s_2$ which are varied along a $400\times 400$ grid. FSLE values are coded with colors, shown by the color bar. 
Enhanced chaotic behavior appears for hard formulae.
} \label{fig:FSLE}
\end{figure}

\paragraph{F. Finite Size Lyapunov Exponents (FSLE).} 
We have
employed the FSLE method \cite{aurell} from nonlinear dynamics theory 
to provide a distributed measure of chaos in our system. 
The FSLE describes the local average 
strength of exponential separation of trajectories and it has been extensively 
used to analyze turbulent flows and processes, including in the atmosphere \cite{koh}
and oceans \cite{ovidio}. The FSLE in a point $\boldsymbol{s}$ is 
given by $\phi(\boldsymbol{s},\varepsilon_0,\varepsilon) = 
\langle \tau^{-1} \ln \varepsilon/\varepsilon_0\rangle $, 
where $\varepsilon_0 = | \boldsymbol{s} - \boldsymbol{s'}|$ is a small initial 
separation of two points with $\boldsymbol{s'}$ chosen in a random direction, 
and $\tau$ is the time needed for the  separation of the corresponding trajectories 
started from these two points to reach the given separation $\varepsilon$. 
The average $\langle \cdot \rangle$ is over the random directions of $\boldsymbol{s'}$. 
In Fig \ref{fig:FSLE} $\varepsilon=30\varepsilon_0$ and $\phi$ is averaged 
for $50$ different randomly oriented initial separations.

\paragraph{G. Random $k$-XORSAT.} In $k$-XORSAT there are given $M$ parity check
equations (constraints) \cite{xors,xorp,MRZ03}, each involving $k$ specified Boolean variables and a parity bit $y_m=\{0,1\}$:
\begin{equation}
x_{i^m_1} + \ldots + x_{i^m_k} \equiv y_m (\mbox{mod 2}),\;\;\;m=1,..,M. \label{XOR}
\end{equation} 
The $k$ distinct variables present in a parity check equation are chosen 
uniformly at random from the set of $x_{1},\ldots,x_{N}$ variables. The goal is to assign
Boolean, $\{0,1\}$ values to the variables such that all parity checks are satisfied. 
As shown in propositional calculus, any propositional formula can be transformed into
its conjugate normal form (CNF), and hence boolean decision problems can all be cast into
a $k$-SAT problem.
Once $k$-XORSAT is transformed into $k$-CNF (Conjugate Normal Form), it becomes
a $k$-SAT problem over a specific ensemble of clauses, hence our dynamical system (1)
can be used to solve $k$-XORSAT (or any boolean decision problem). 
For $k=3$ we use four 3-SAT clauses to encode one parity check equation.

Random XORSAT  also goes through phase transitions \cite{xorp,MRZ03} when increasing the 
constraint density $\gamma=M/N$. The dynamical phase transition takes place
at $\gamma_d$ (for $3$-XORSAT $\gamma_d=0.8185$), when the unique solution 
cluster existing at $\gamma < \gamma_d$  breaks into an exponentially large 
number of clusters. The second phase transition is the SAT/UNSAT transition at 
$\gamma_c > \gamma_d$, with $\gamma_c=0.9179$ for $3$-XORSAT \cite{MRZ03}.
\begin{figure}[htbp]
\vspace*{-0.4cm}
\centerline{\includegraphics[width=5.5in]{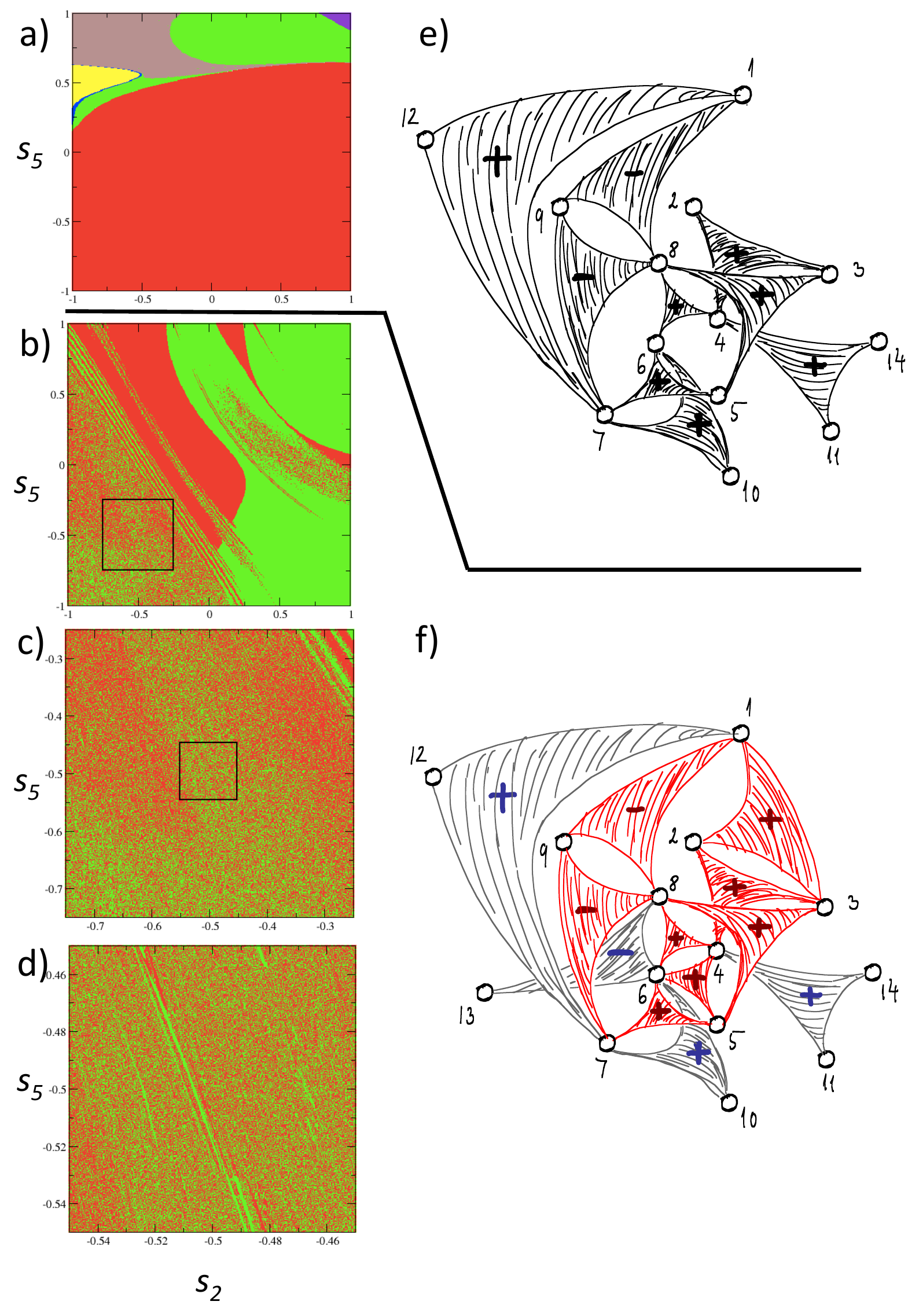}} 
\vspace*{-0.5cm}
\caption{{\bf 3-XORSAT}. 
a) basins of attraction to solutions for a formula
given by the hypergraph in e), which has no hyperloops (core). b) basins of attraction 
to solutions for the formula in f), which has a core (shown in red).
c) is the magnification of the small black rectangle from b), and d) is a magnification of a small
black rectangle from c). In the hypergraphs e) and f) the ``triangles" represent 
hyperedges/parity checks ($k=3$) and the signs correspond to $y_j$-s. } \label{3xorbasins}
\end{figure}
Constraint satisfaction formulae can be represented as hypergraphs with
nodes representing the variables and with hyperlinks representing the constraints
connecting the variables present in them. To find the various transitions for $k$-XORSAT,
however, it is actually better not to bring it into CNF form (we do that for simulations 
with our system (\ref{sdyn}),(\ref{adyn}). Using this representation, it was 
proven \cite{MRZ03} that $\gamma_d$ corresponds to 
the constraint density where the hyperloops (in this non-CNF form hypergraph)  
appear with non-zero statistical weight in the limit of $N \to \infty$. 
As chaotic behavior in our dynamical system (1) appears already at one of the 
smallest (finite) hyperloop motifs, it therefore also appears at exactly 
the same $\gamma_d$ that M\'ezard et.al. calculated \cite{MRZ03}.
Fig \ref{3xorbasins}e) shows the full hypergraph for a small specific instance of 
3-XORSAT without a {\em core} at $\gamma = 9/14 = 0.64$ (after performing the leaf-removal
algorithm described in \cite{MRZ03} the remaining hyperloops form the core of 
the XORSAT instance). As the graph in Fig \ref{3xorbasins}e) has no core (no hyperloops), 
the corresponding basin boundary is indeed smooth, i.e., there is no (transient) 
chaos, also illustrated in Fig \ref{3xorbasins}a). However, once we add more constraints 
(for example the specific ones shown in Fig \ref{3xorbasins}f)), giving 
$\gamma = 12/14 = 0.85$  the core appears (shown in red in 
Fig \ref{3xorbasins}f)), and the basin boundaries become fractal, see Fig \ref{3xorbasins}b)-d).

\paragraph{H. Polynomial continuous-time complexity for a fixed number of formulae.} 
We now show that Eq (2) of the main text  implies the stronger result of having polynomial 
continuous - time 
complexity even for leaving a {\em fixed number} (not fraction!) of formulae unsolved in the limit $N \to \infty$. For a given $N$ and $M = \alpha N$
there are a total of $$\Theta^{(k)}_{\alpha}(N) = {2^k {N \choose k} \choose \alpha N}$$ 
$k$-SAT 
formulae. If $c$ is a small integer constant (independent of $N$), setting 
$p = c / \Theta^{(k)}_{\alpha}(N)$ in (2) of the main text yields
$$t(p,N) = b^{-1} N^{\beta} \left[ \ln(r/c) + \ln \Theta^{(k)}_{\alpha}(N) \right].$$ 
However, for fixed $\alpha$, 
$\ln \Theta^{(k)}_{\alpha}(N)  = \alpha (k-1)N \ln N + {\cal O}(\ln N)$ as $N \to \infty$, which implies
that $$t(p,N) \sim N^{\beta+1} \ln N,$$ i.e., still showing polynomial CT complexity.  
Thus, assuming that (2) holds for all $N$, the fact that only a constant
number of problems may not be solved in polynomial time for $N \to \infty$ indicates that
the algorithm runs in polynomial continuous-time on {\em almost all} hard $k$-SAT instances 
(the probability
to find an instance not solvable in polynomial time by this solver is 
zero in the $N\to \infty$ limit). 

\begin{figure}[htbp]
\vspace*{-0.3cm}
\centerline{\includegraphics[width=2.0in]{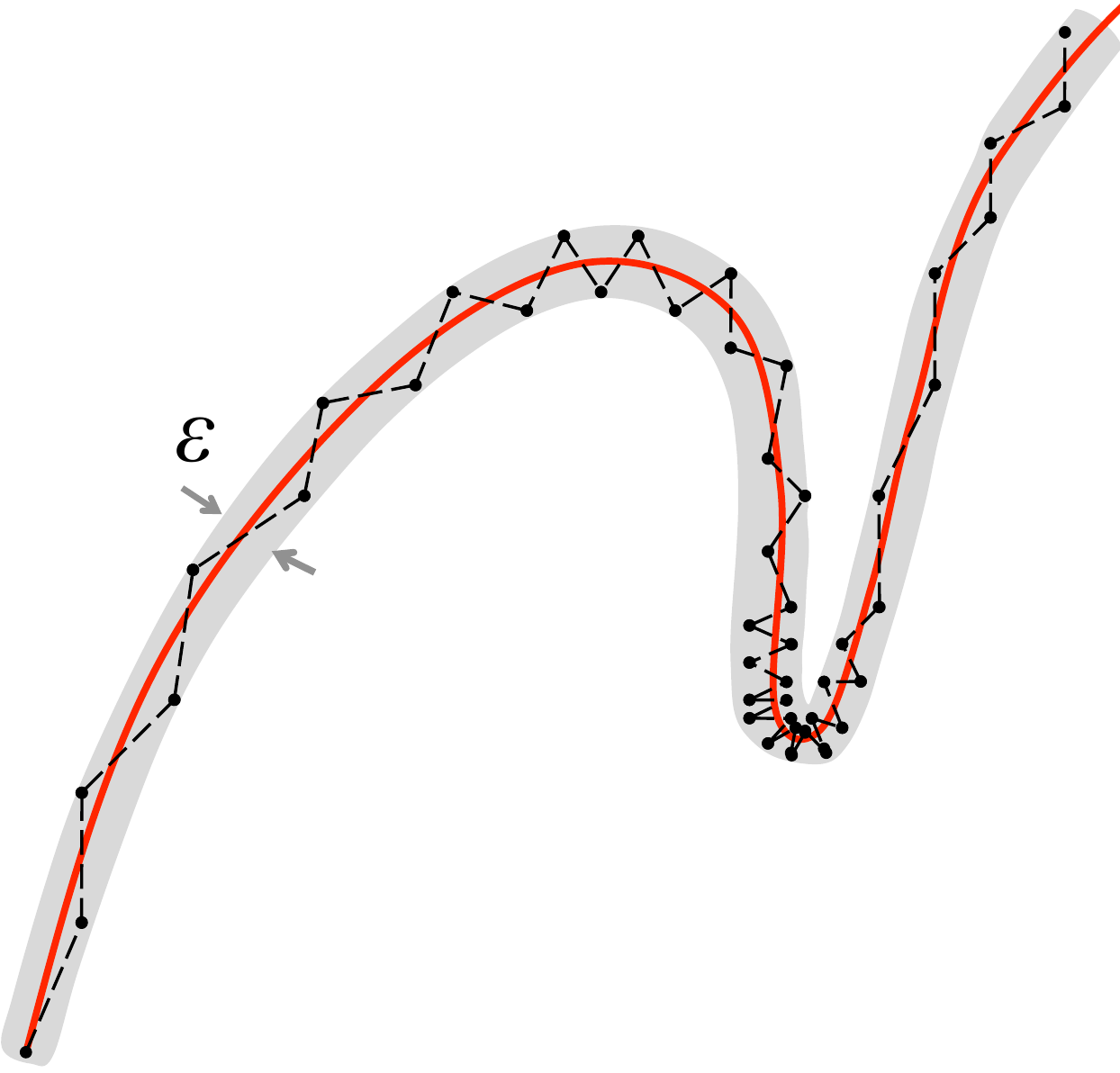}} 
\vspace*{-0.3cm}
\caption{{\bf Approximating the continuous-time trajectory.} 
The RK algorithm must compute
a large number of discrete points in high curvature regions in 
order to stay within a prescribed precision $\varepsilon$.
} \label{fig:Scikkcakk}
\end{figure}

\begin{figure}[htbp]
\centerline{\includegraphics[width=6.2in, height = 2.4in]{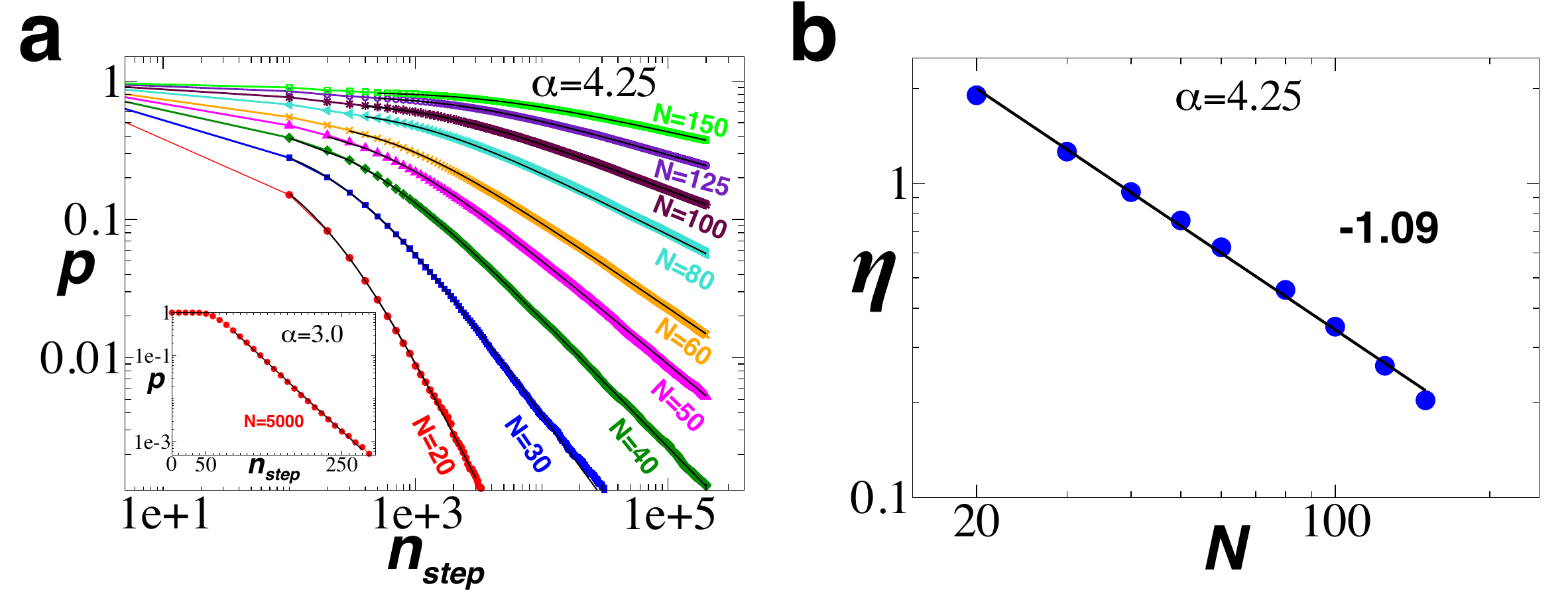}}
\caption{{\bf Discrete time complexity.} {\bf a:}  Fraction of problems 
$p(n_{step})$  left unsolved after $n_{step}$ discretization 
steps, at $\alpha=4.25$, $k=3$
for different system sizes $N=20$, $30$, $40$, $50$, $60$, 
$80$, $100$, $125$, $150$ (different colors). 
Averages were done over $10^5$ instances for each $N$, except for $N=150$ where 
$3 \times 10^4$ instances were used. The black continuous lines
are $p(n_{step}) = u(v + n_{step})^{-\eta}$. 
The inset (log-lin plot) shows $p(n_{step})$ from $10^5$ instances 
with $N=5000$ at $\alpha=3$. In this easy-SAT region
$p(n_{step})$ shows a similar exponential decay as $p(t)$ shown in Fig3a, main text. 
{\bf b:} The exponent $\eta$ follows: $\eta(N)=dN^{-\delta}$, $\delta\simeq 1.09$.}
\label{fig:disc-cont}
\end{figure}

\paragraph{I. Computational complexity in the number of discrete steps.}
Since the numerical integration happens on a Turing machine, 
the true continuous trajectory is  being approximated by the Runge-Kutta (RK) algorithm
\cite{numrecipes} 
with sufficiently many 
discrete points, lying close to it, within a tube of preset diameter $\varepsilon$ (see Fig.
\ref{fig:Scikkcakk}). 
When we monitor the fraction of formulae
left unsolved $p$ as function of the {\em number of discretization steps} $n_{step}$ taken 
by the algorithm for hard SAT formulae (from the frozen phase), we find that
$p(n_{step})$ has a {\em power-law} decay, well approximated by 
$p(n_{step})=  u(v+n_{step})^{-\eta}$, see 
Fig. \ref{fig:disc-cont}a. 
The exponent $\eta$ also has
 a power-law $N$-dependence (see Fig. \ref{fig:disc-cont}b): $\eta(N) = d N^{-\delta}$ 
 with $\delta \simeq 1.09 \simeq 1$ (the $N$-dependence of $u$ and $v$ are weak). 
 This implies an {\em exponential behavior} for the number of time steps $n_{step}(p,N)$
 needed to miss solving only a $p$-th fraction of the formulae: 
 \begin{equation}
 n_{step}(p,N) = e^{N^{\delta} \frac{1}{d} \ln \frac{u}{p}} - v,
 \end{equation}
showing exponential time-complexity for the discretized algorithm ran by a
digital computer (Turing machine). For easy formulae, however, (such as those drawn at random
for $\alpha = 3$),  $p(n_{step})$
has an exponential decay, just as $p(t)$, implying polynomial complexity for the
discretized algorithm as well, see the inset of Fig \ref{fig:disc-cont}a. 

\begin{figure}[htbp] 
\centerline{\includegraphics[width=5in]{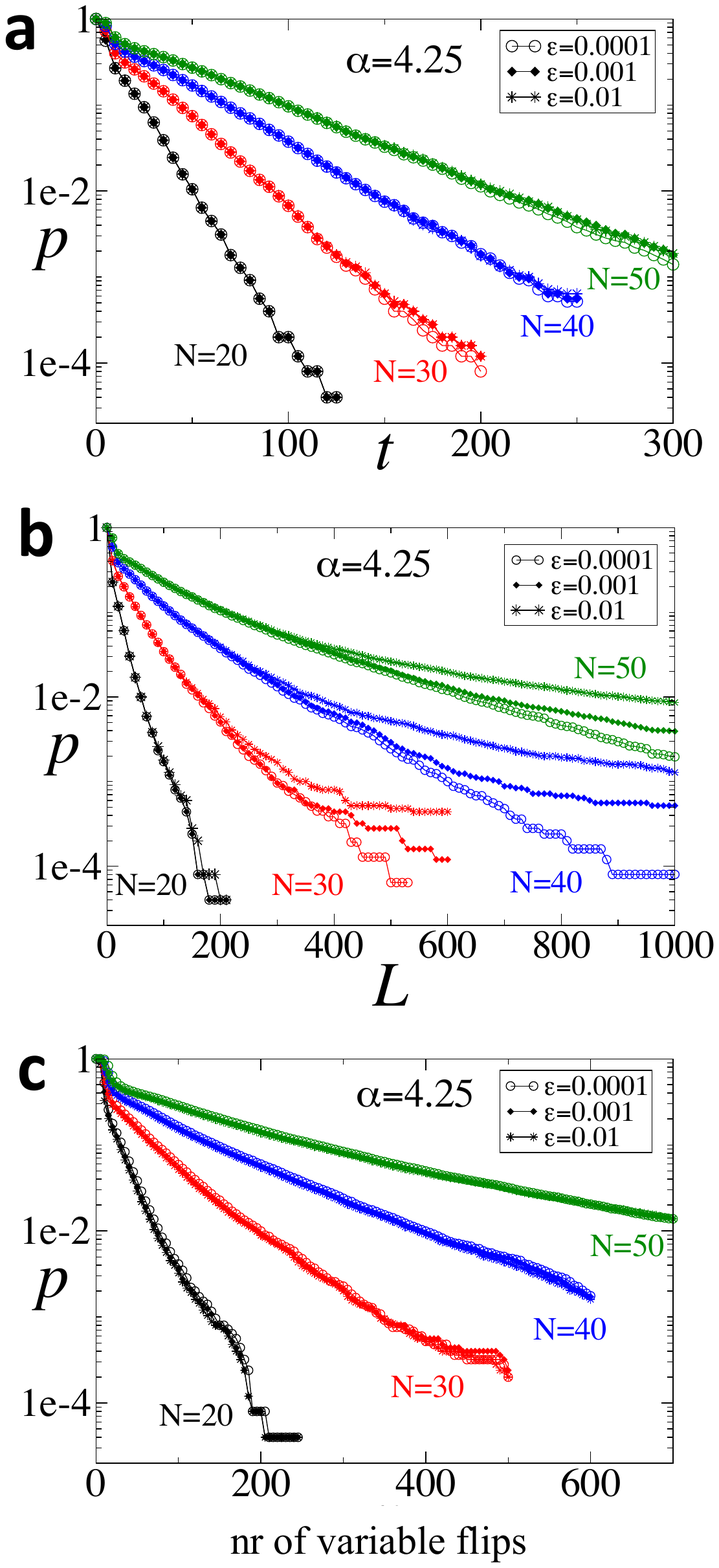} }
\vspace*{-0.3cm}
\caption{ {\bf Sensitivity to discretization error.} {\bf a:} Fraction of problems $p(t)$ left unsolved by continuous-time 
$t$, at $\alpha=4.25$, $k=3$ for system sizes $N=20,30,40,50$ (different colors) 
and for different error parameters of the adaptive Runge-Kutta method \cite{numrecipes}: 
$\epsilon=0.0001, 0.001, 0.01$. Here $\epsilon$ is the maximal relative error allowed 
during the RK integration. Statistics was done on $2.5 \times 10^4$ instances for each $N$. 
{\bf b:}  The fraction of problems $p(L)$ left unsolved by trajectories of length at 
most $L$ (measured in ${\cal H}_N$), 
for the same problems as in {\bf a}. }
\label{fig:error}
\end{figure}

\paragraph{J. Dependence on discretization error.} The continuous-time variable $t$ 
and the decay of $p$ as function of $t$ is only weakly dependent on the 
discretization error $\varepsilon$ by the RK solver, with a slight shift towards 
even cleaner exponentials when lowering $\varepsilon$, see Fig. \ref{fig:error}a.
 
As stated in the main text, the polynomial CT complexity is not the consequence of
a log-transformation on the time variable $t$. The scaling of the {\em length} $L$
of the continuous-time trajectory measured from the initial point until it finds a
solution also scales polynomially with $N$. However, this measure is more sensitively
affected by the discretization error $\varepsilon$.  This is because, as illustrated in 
Fig. \ref{fig:Scikkcakk}, the length computed as the sum of the lengths of the straight segments 
between discretization steps {\em overestimates}  the analog trajectory length 
(continuous red line). Fig. \ref{fig:error}b shows the
fraction of formulae $p$ left unsolved by trajectories of length not longer than 
$L$ (measured in ${\cal H}_N$),
as function of $L$ for different discretization errors $\varepsilon$. The convergence
to clean exponentials is evident as $\varepsilon$ is lowered.

\begin{figure}[htbp]
\centerline{\includegraphics[width=6.2in]{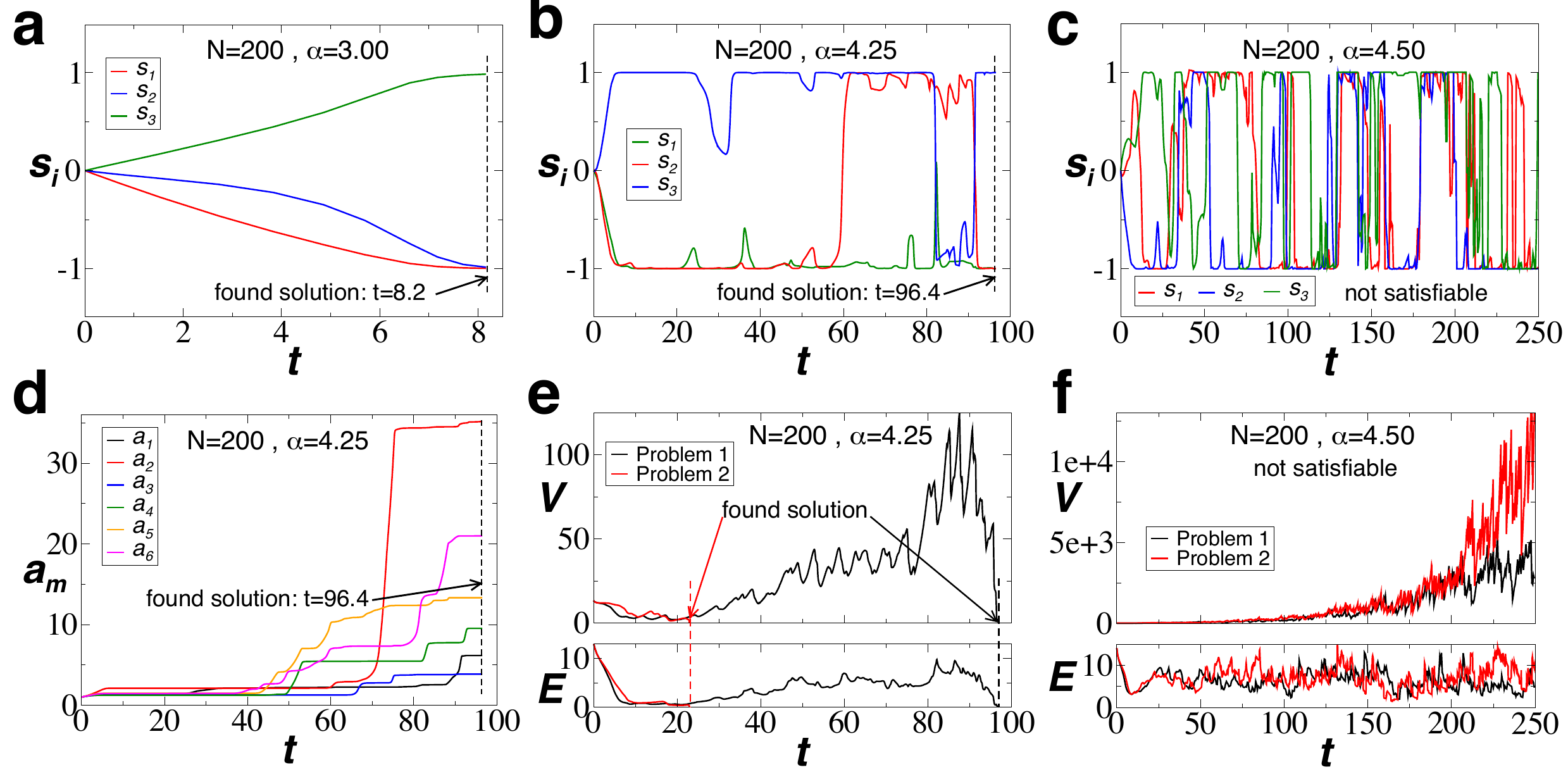}} 
\caption{{\bf Time evolution of variables and energy functions.} Time evolution of three variables $s_i(t)$ 
(different colors) for a formula with $k=3$, $N=200$ and {\bf a:}  $\alpha=3$, 
characterized by rapid and straightforward convergence; in {\bf b:} at $\alpha=4.25$, 
presenting much longer, and chaotic trajectories, and {\bf c:} for  an unsatisfiable formula at $\alpha=4.5$. 
{\bf d:} Time series of six different auxiliary variables $a_m(t)$ shown with different colors for the same 
problem as in {\bf b}. Time series of  $E({\bf s})$ and $V({\bf s},{\bf a})$ as function of the continuous time $t$, for two different problems (red and black) with $k=3$, $N=200$ at {\bf e:} $\alpha=4.25$ and {\bf f:} for two unsatisfiable formulae 
at $\alpha=4.5$.
} \label{fig:trajectories}
\end{figure}

\begin{figure}[htbp]
\vspace*{-0.5cm}
\begin{center}
\includegraphics[width=6.0in]{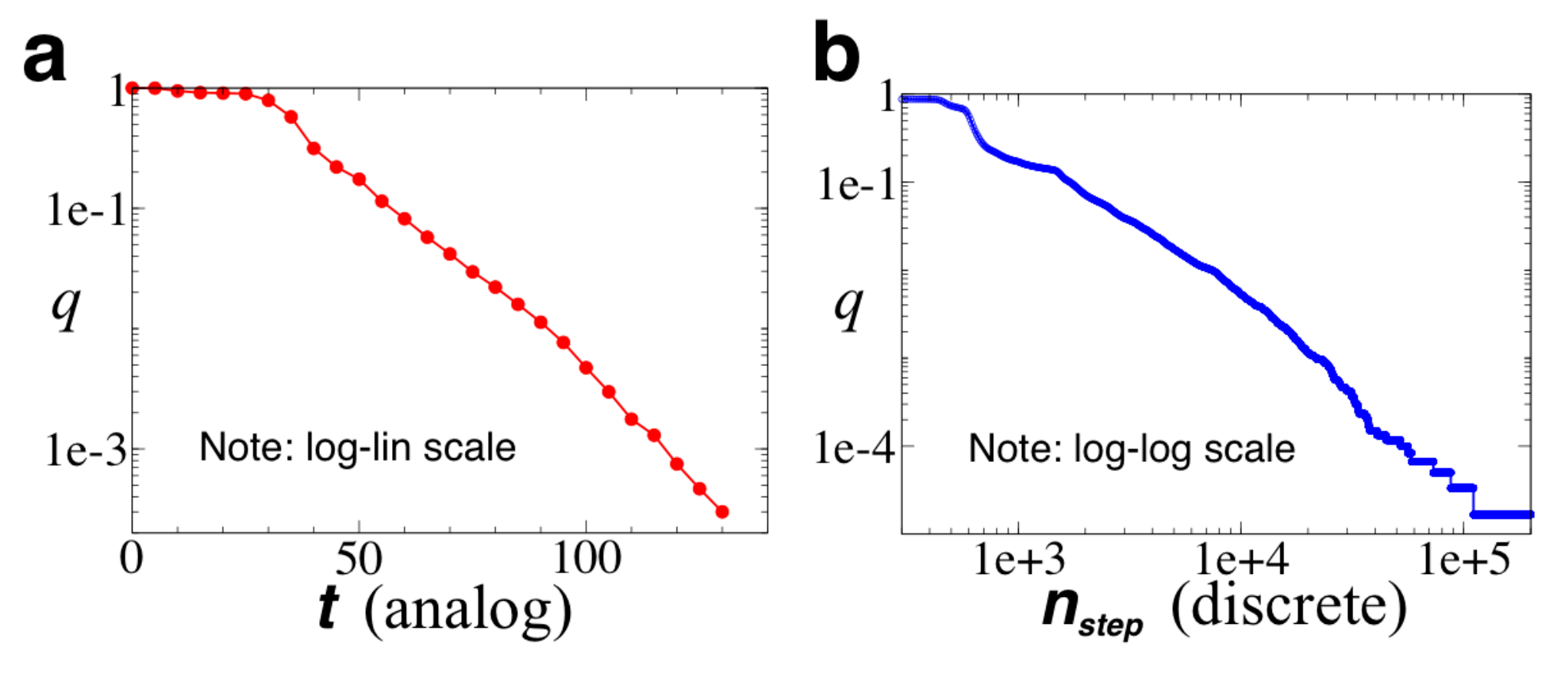}
\label{basin}  \vspace*{-0.5cm}
\caption{{\bf Exponential decay for a single instance. } For single 3-SAT instance with $N=40$, 
at $\alpha=4.25$ we start the dynamics from
$60000$ different random initial conditions. {\bf a:} Similarly to Fig 3a of the main text the fraction of trajectories $q(t)$, 
which did not find the solution by analog-time $t$ shows an exponential decay as function of $t$. {\bf b:} The fraction of trajectories, $q(n_{step})$
still searching for the solution after $n_{step}$ discretization steps, however, shows a power-law behavior. }
\end{center}\vspace*{-0.5cm}
\end{figure}
\begin{figure}[htbp]
\vspace*{-0.5cm}
\begin{center}
\includegraphics[width=5.0in]{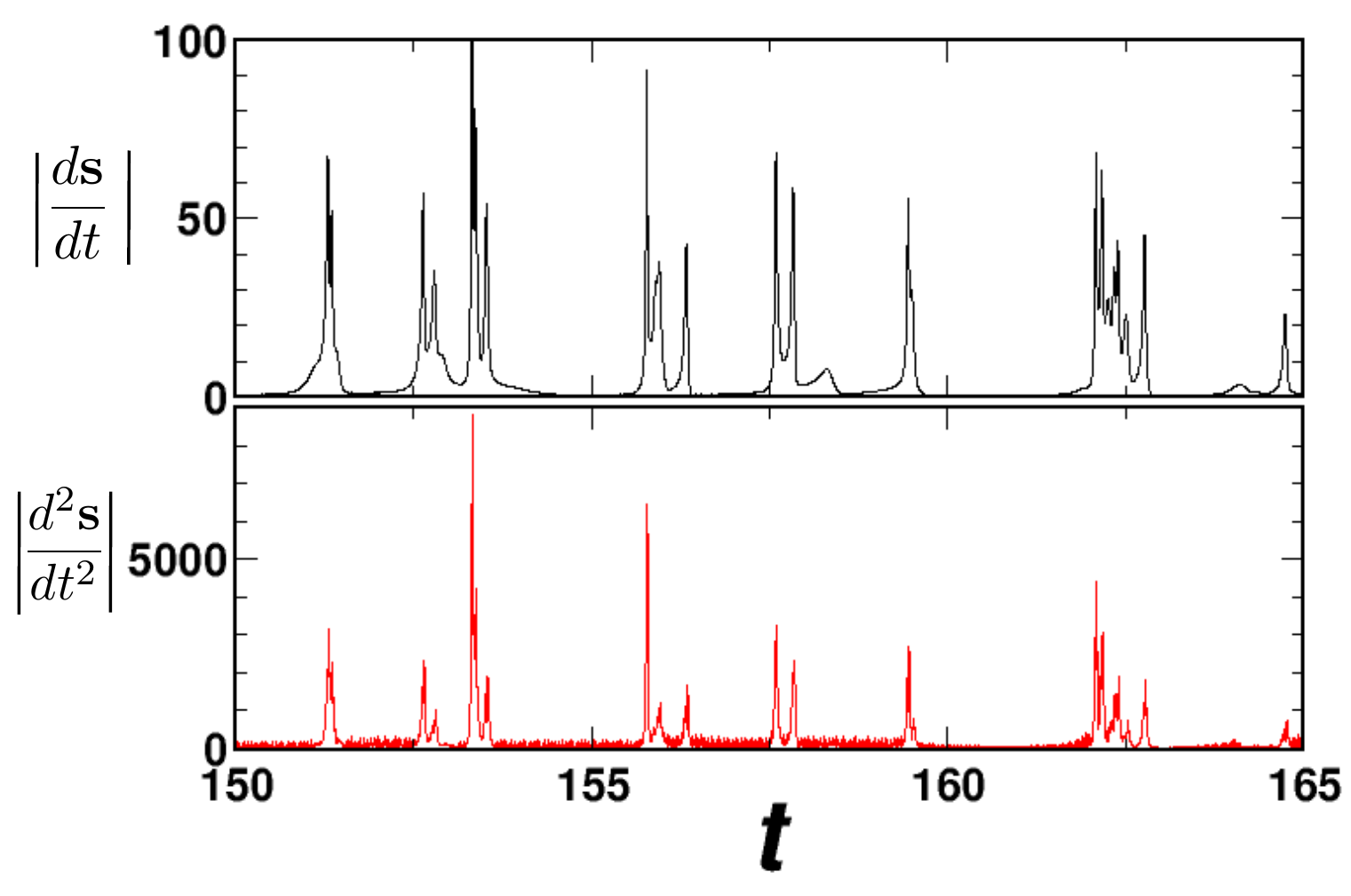}\label{basin}  \vspace*{-0.2cm}
\caption{{\bf Intermittent behaviour}. {\bf a:} The velocity $|d\bm{s}/dt|$ and the acceleration $|d^2\bm{s}/dt^2|$
 along the trajectories in the hard-SAT phase ($N=100$, $\alpha=4.25$, $3$-SAT) show intermittency
as function of time similar to turbulent flows \cite{MS-1991}. {\bf b:} Same as {\bf(a)} on log-linear scale.
}\label{interm}
\end{center}
\end{figure}

\paragraph{K. Analogy with fluid turbulence}
Continuous-time processes are common in nature, from fluid flows to 
information processing in the brain; even our perception of time is arguably of analog nature. 
While transient chaos can appear in many dynamical systems, 
long chaotic transients typically occur in a parameter region preceding 
the permanently chaotic regime. 
If the average lifetime of transients in a dynamical system depends on an extensive 
parameter of the system, supertransients (or superpersistent chaotic transients) 
may appear  \cite{Lai-Tel-PR08,Lai-Tel-2011}. 
The only experiments confirming the existence of such supertransients 
are those done in long-pipe flow experiments.  Here, turbulent/chaotic behavior appears 
before the flow enters its laminar phase (parabolic velocity profile), which is its only asymptotic attractor \cite{Hof-Nature,Hof-PRL}.  
One can think of fluid turbulence in this case as nature's search for equilibrium, practically solving a hard global optimization problem. Fig \ref{interm} perhaps takes this analogy further: it shows the fluctuations of the instantaneous velocity and acceleration for a typical trajectory for a hard formula from the frozen region as function of time, revealing intermittent behavior, typically found in turbulence, see Fig 1 of the paper by Meneveau and Sreenivasan \cite{MS-1991}.

\newpage


\end{document}